\documentclass[12pt]{article}
\pdfoutput=1
\usepackage{tikz}
\usepackage{a4wide,epsfig,psfrag,amsmath,amssymb,cite,scalefnt}
\usepackage{color}
\usepackage{amsmath,comment,braket}
\usepackage{placeins}
\usepackage{slashed}

\graphicspath{ {figs/} }

\definecolor{redmar}{rgb}{.8,0,0}

\allowdisplaybreaks

\newcommand{\gsim}{\;\rlap{\lower 3.5 pt \hbox{$\mathchar \sim$}} \raise 1pt
 \hbox {$>$}\;}
\newcommand{\lsim}{\;\rlap{\lower 3.5 pt \hbox{$\mathchar \sim$}} \raise 1pt
 \hbox {$<$}\;}

\newcommand{\rd}{\mathrm d}
\newcommand{\f}[2]{\frac{#1}{#2}}
\newcommand{\lb}{\left(}
\newcommand{\rb}{\right)}

\newcommand{\lbe}{\left[}
\newcommand{\rbe}{\right]}
\newcommand{\ri}{\mathrm i}
\newcommand{\calI}{\mathcal{I}}
\newcommand{\calS}{\mathcal{S}}
\newcommand{\calT}{\mathcal{T}}
\newcommand{\calP}{\mathcal{P}}
\newcommand{\calO}{\mathcal{O}}
\newcommand{\ord}{\mathcal O}
\newcommand{\MB}{\scriptscriptstyle\mathrm{(MB)}}

\begin{document}

\title{\vskip-3cm{\baselineskip14pt
    \begin{flushleft}
      \normalsize P3H-22-063\\
      \normalsize TTP22-041 \\
      \normalsize TU-1160 \\
  \end{flushleft}}
  \vskip1.5cm
  Higgs boson contribution to the leading two-loop Yukawa corrections to $gg\to HH$
}

\author{ Joshua Davies$^{a}$, Go Mishima$^{b}$, Kay Sch\"onwald$^{c}$,\\
  Matthias Steinhauser$^{c}$, Hantian Zhang$^{c}$
  \\[1mm]
  {\small\it $^a$ Department of Physics and Astronomy, University of Sussex,
    Brighton BN1 9QH, UK}
  \\[1mm]
  {\small\it $^b$ Department of Physics, Tohoku University, Sendai, 980-8578
    Japan}
  \\[1mm]
  {\small\it $^c$Institut f{\"u}r Theoretische Teilchenphysik, Karlsruhe Institute of Technology (KIT)}\\
  {\small\it Wolfgang-Gaede Stra\ss{}e 1, 76128 Karlsruhe, Germany} }
  
\date{}

\maketitle

\thispagestyle{empty}

\begin{abstract}

  We analytically compute two-loop Yukawa corrections to Higgs boson pair
  production in the high-energy limit.  Such corrections are generated by an
  exchange of a Higgs boson between the virtual top quark lines.  We propose two
  approaches to obtain expansions of the massive two-loop box integrals and
  show that precise results are obtained for transverse momenta of the Higgs bosons
  above about 150~GeV.  We discuss in detail the computation of all 140
    master integrals and present analytic results.

\end{abstract}

\thispagestyle{empty}

\sloppy


\newpage


\section{Introduction}

Higgs boson pair production is a promising process which can provide experimental
information about the Higgs boson self coupling
(see, e.g., Ref.~\cite{LHCHiggsCrossSectionWorkingGroup:2016ypw}).
It is thus
important to provide precise theoretical predictions of this process.  The dominant
contribution to Higgs boson pair production comes from gluon fusion,
mediated by a top quark loop.  There are a number of works in the literature in which QCD
corrections to $gg\to HH$ have been considered. The NLO QCD corrections are
known exactly~\cite{Borowka:2016ehy,Borowka:2016ypz,Baglio:2018lrj}, however,
the numerical approach is quite computationally demanding. In practice it is therefore
advantageous to construct approximations based on several expansions, valid in
different regions of phase
space~\cite{Dawson:1998py,Grigo:2013rya,Degrassi:2016vss,Davies:2018ood,Davies:2018qvx,Bonciani:2018omm,Grober:2017uho,Xu:2018eos,Wang:2020nnr}.
A subsequent combination of the numerical approach with these expansions leads to
fast and precise results which cover the whole phase
space~\cite{Davies:2019dfy,Bellafronte:2022jmo}.  At
NNLO~\cite{deFlorian:2013jea,deFlorian:2013uza,Grigo:2014jma,Grigo:2015dia,Davies:2019xzc,Davies:2021kex}
and
N$^3$LO~\cite{Spira:2016zna,Gerlach:2018hen,Banerjee:2018lfq,Chen:2019lzz,Chen:2019fhs}
only the large-$m_t$ expansion has been considered.  The to date most precise
predictions have been obtained in Ref.~\cite{Grazzini:2018bsd} where a NNLO
approximation has been constructed, based partly on exact and partly on
large-$m_t$ results.

\begin{figure}[t]
  \centering
  \includegraphics[width=0.5\textwidth]{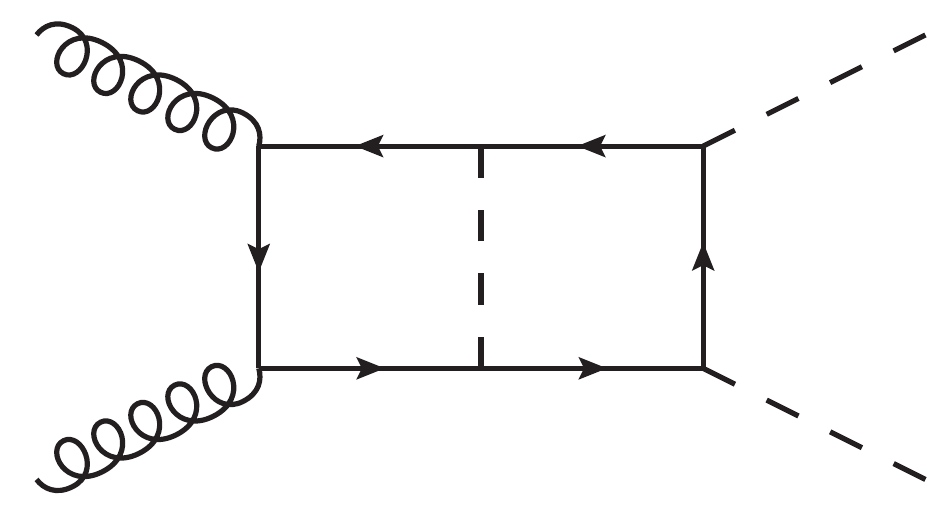}
  \caption{\label{fig::diags1}Sample Feynman diagram with a Higgs boson
    exchange in the top quark loop. Straight, dashed and curly lines
    represent top quarks, Higgs bosons and gluons, respectively.}
\end{figure}

Electroweak corrections are expected to be of the order of a few percent
and thus they should be included in the theoretical description.
In the Standard Model there are several couplings (gauge, Yukawa, Higgs boson self coupling)
which are of different nature and can be treated separately.  In this paper
we take a first step towards the electroweak corrections and compute top quark
Yukawa corrections originating from Higgs boson exchange in the top
quark loop.  More precisely, we consider diagrams like the one shown in
Fig.~\ref{fig::diags1}. For this subclass only planar diagrams contribute and
thus only planar integral families have to be considered.

Note that in the $R_\xi$ gauge there are also other Yukawa corrections from the
exchange of neutral and charged Goldstone bosons. They are not considered in
this paper.  Rather we concentrate on corrections with a virtual Higgs boson.

In the case of QCD corrections the top quark is the only massive particle in
the loop. As additional scales, one has the Mandelstam variables $s$ and $t$
and the Higgs boson mass from the final-state particles.  Electroweak
corrections introduce additional masses in the propagators of the loop
integrals, which increases the complexity significantly.

\begin{figure}[t]
  \centering
  \begin{tabular}{cccc}
    \includegraphics[width=0.2\textwidth]{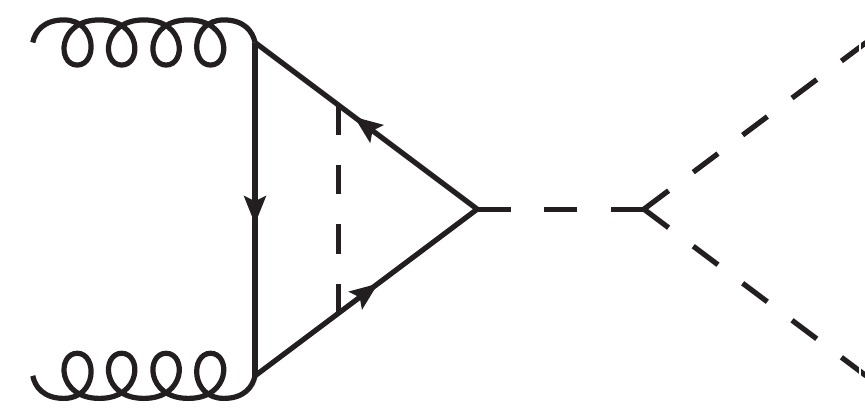} &
    \includegraphics[width=0.2\textwidth]{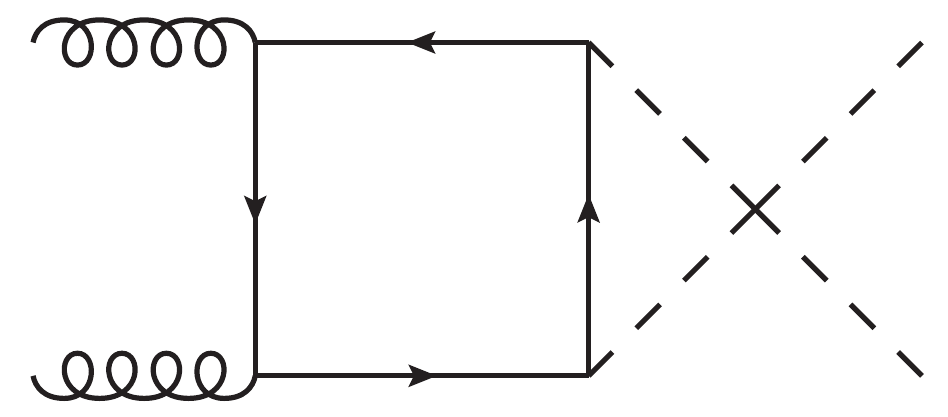} &
    \includegraphics[width=0.2\textwidth]{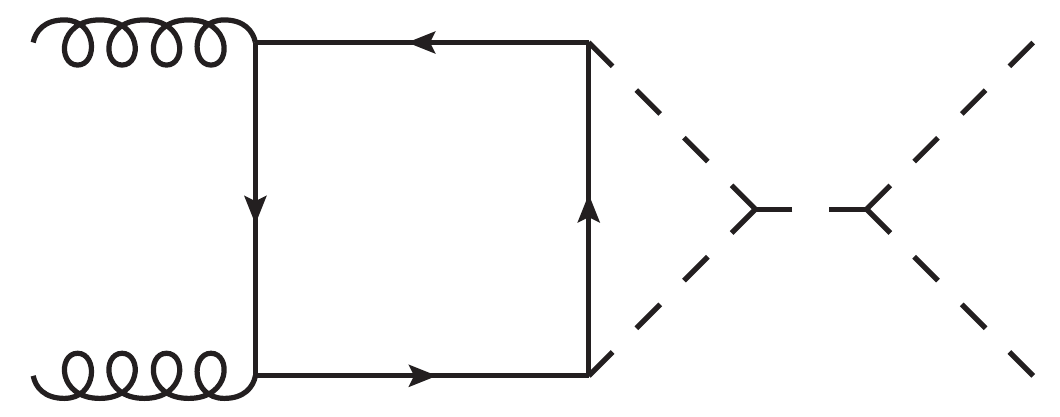} &
    \includegraphics[width=0.2\textwidth]{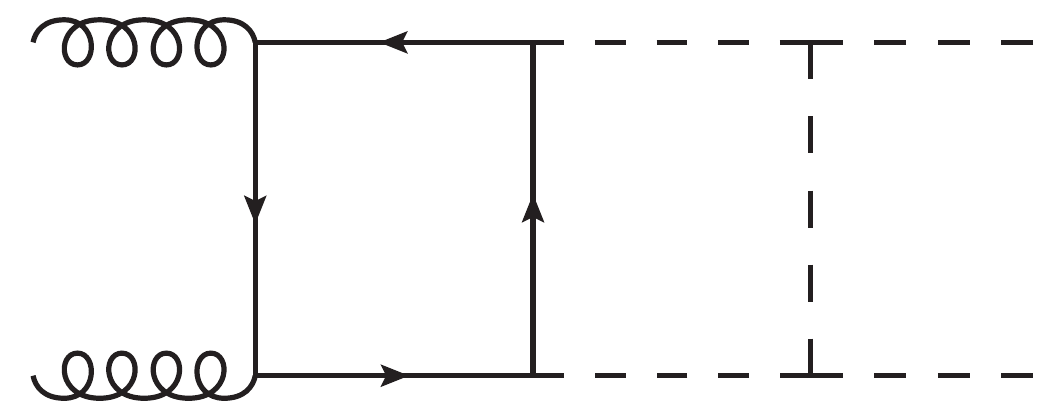} \\
    (a) & (b) & (c) & (d) \\
    \includegraphics[width=0.2\textwidth]{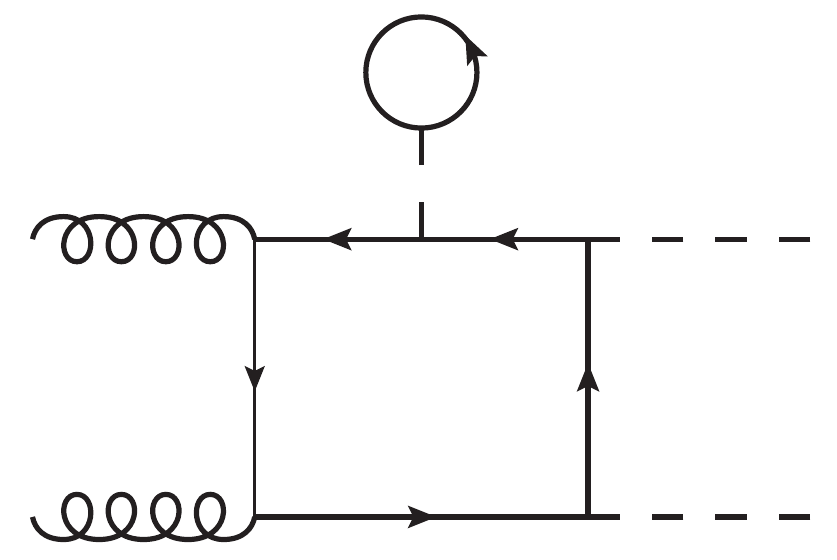} &
    \includegraphics[width=0.2\textwidth]{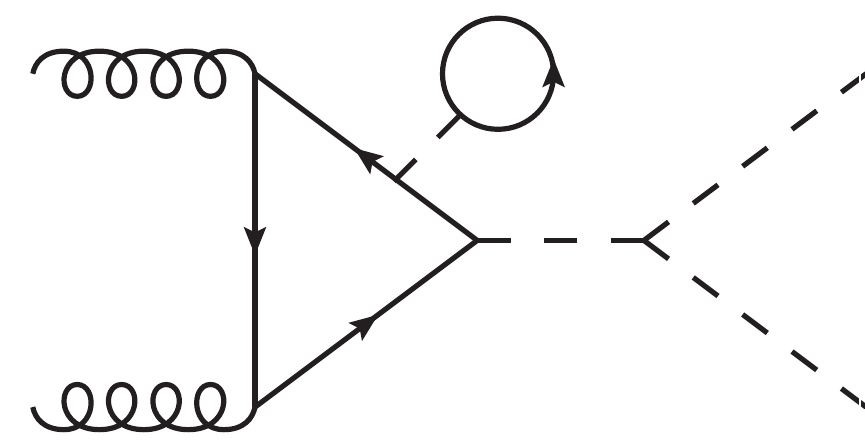} &
    \includegraphics[width=0.2\textwidth]{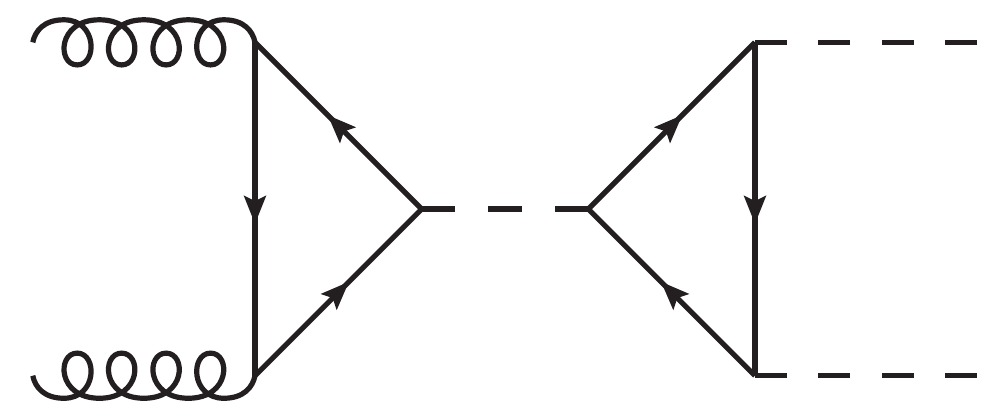} &
    \includegraphics[width=0.2\textwidth]{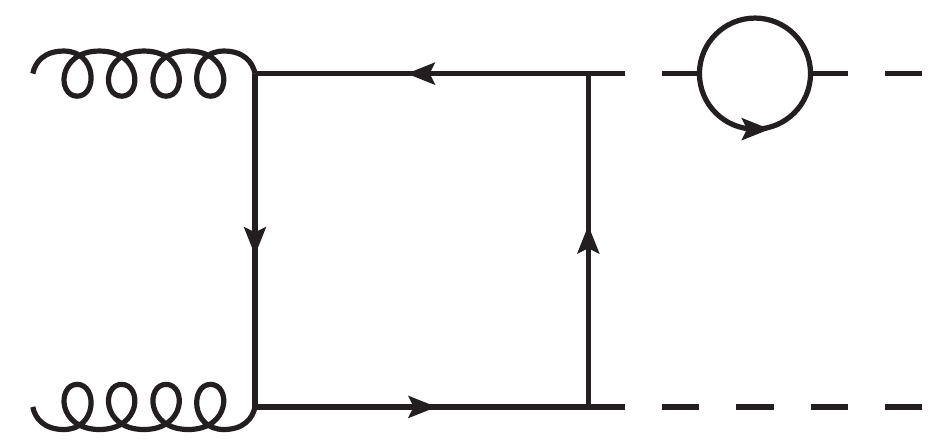} \\
    (e) & (f) & (g) & (h) \\
  \end{tabular}
  \caption{\label{fig::diags2}Diagrams with Higgs boson self coupling and
    one-particle reducible diagrams. These classes of diagrams are not
    considered in this paper.}
\end{figure}

There are further classes of diagrams with a Higgs boson exchange.  In
contrast to the diagram in Fig.~\ref{fig::diags1} they either involve Higgs
boson self couplings (see Fig.~\ref{fig::diags2}(a)-(d)) or are one-particle
reducible (see Fig.~\ref{fig::diags2}(e)-(h)).  The results for the triangle
diagrams in Fig.~\ref{fig::diags2}(a) can be obtained from the integral
families discussed in this paper.  Note that the diagram classes (b), (c) and
(d) also involve non-planar contributions. Diagrams (e)-(h) are one-particle
reducible and factorize into a product of one-loop integrals.

The master integrals which are computed in this paper are
sufficient to compute the contributions from Figs.~\ref{fig::diags2}(a), (e),
(f), (g) and (h). However, in this paper we concentrate on the
two-loop box contribution of Fig.~\ref{fig::diags1} and pursue the
following goals:
\begin{itemize}
\item Develop a method to obtain high-energy approximations of two-loop
  four-point integrals where two different masses are present inside the loops.
  
\item Provide details of the analytic computation of the master integrals
  which appear in the subclass of diagrams considered in this paper.

\item 
  Provide explicit analytic results for the master integrals in the high-energy limit.
\end{itemize}

The remainder of the paper is organized as follows:~in the next section
we introduce our notation and in Section~\ref{sec:asyexp} we outline
the expansions which we apply to the Feynman
diagrams. In Section~\ref{sec::amplitude} details of the computation of the
amplitudes in terms of master integrals are provided.
In Section~\ref{sec::full-mt-MI} we provide a detailed description
of the computation of the master integrals and numerical results
of the form factors are are given in Section~\ref{sec::FF}.
We conclude in Section~\ref{sec::concl}. In the appendix we present results
for three-dimensional Mellin-Barnes integrals
which enter our result.


\section{Notation}

The Mandelstam variables for the amplitude $g(q_1)g(q_2)\to H(q_3)H(q_4)$, with all
momenta ($q_i$) defined to be incoming, are given by
\begin{eqnarray}
  {s}=(q_1+q_2)^2\,,\qquad {t}=(q_1+q_3)^2\,,\qquad {u}=(q_2+q_3)^2\,,
  \label{eq::stu}
\end{eqnarray}
with
\begin{eqnarray}
  q_1^2=q_2^2=0\,,\qquad  q_3^2=q_4^2=m_H^{2}\,,\qquad
  {s}+{t}+{u}=2m_H^{2}\,.
  \label{eq::q_i^2}
\end{eqnarray}
It is convenient to introduce the scattering angle $\theta$ and the transverse
momentum of the Higgs bosons in the center-of-mass frame, which are given by
\begin{eqnarray}
\label{eqn:thetadef}
  p_T^2 &=& \frac{{u}{t}-m_H^4}{{s}}\,,
          \nonumber\\
  t &=& m_H^2 - \frac{s}{2}\left(1-\cos\theta
        \sqrt{1-\frac{4m_H^2}{s}}\right)\,.
\end{eqnarray}

Due to Lorentz and gauge invariance it is possible to define two scalar matrix
elements ${\cal M}_1$ and ${\cal M}_2$ as
\begin{eqnarray}
  {\cal M}^{ab} &=& 
  \varepsilon_{1,\mu}\varepsilon_{2,\nu}
  {\cal M}^{\mu\nu,ab}
  \,\,=\,\,
  \varepsilon_{1,\mu}\varepsilon_{2,\nu}
  \delta^{ab}
  \left( {\cal M}_1 A_1^{\mu\nu} + {\cal M}_2 A_2^{\mu\nu} \right)
  \,,
\end{eqnarray}
where $a$ and $b$ are adjoint colour indices and the two Lorentz structures
are given by
\begin{eqnarray}
  A_1^{\mu\nu} &=& g^{\mu\nu} - {\frac{1}{q_{12}}q_1^\nu q_2^\mu
  }\,,\nonumber\\
  A_2^{\mu\nu} &=& g^{\mu\nu}
                   + \frac{1}{p_T^2 q_{12}}\left(
                   q_{33}    q_1^\nu q_2^\mu
                   - 2q_{23} q_1^\nu q_3^\mu
                   - 2q_{13} q_3^\nu q_2^\mu
                   + 2q_{12} q_3^\mu q_3^\nu \right)\,,
\end{eqnarray}
with $q_{ij} = q_i\cdot q_j$.
The Feynman diagrams involving the Higgs boson self coupling only
contribute to $A_1^{\mu\nu}$ and thus, it is convenient to decompose
${\cal M}_1$ and ${\cal M}_2$ into ``triangle'' and ``box'' form factors
\begin{eqnarray}
  {\cal M}_1 &=& X_0 \, s \, \left(\frac{3 m_H^2}{s-m_H^2} F_{\rm tri} + F_{\rm box1}\right)
                 \,,\nonumber\\
  {\cal M}_2 &=& X_0 \, s \, F_{\rm box2}
                 \,,
                 \label{eq::calM}
\end{eqnarray}
with
\begin{eqnarray}
  X_0 &=& \frac{G_F}{\sqrt{2}} \frac{\alpha_s(\mu)}{2\pi} T \,,
\end{eqnarray}
where $T=1/2$, $\mu$ is the renormalization scale and $G_F$ is the Fermi constant.

We define the perturbative expansion of the form factors as
\begin{eqnarray}
  F &=& F^{(0)} 
        + \frac{\alpha_s(\mu)}{\pi} F^{(1,0)} 
        + \frac{\alpha_t}{\pi} F^{(0,y_t)} 
        + \cdots
  \,,
  \label{eq::F}
\end{eqnarray}
where $\alpha_t$ is given by
\begin{eqnarray}
  \alpha_t &=& \frac{\alpha \, m_t^2}{2 \, s_W^2 m_W^2}\,.
               \label{eq::alphat}
\end{eqnarray}
$\alpha$ is the fine structure constant and $s_W^2 \equiv \sin^2\theta_W$ is the square of
the sine of the weak mixing angle.
Throughout this paper the strong coupling constant is defined with six active
quark flavours.  Note that the form factors are defined such that the one-loop
colour factor $T$ is contained in the prefactor $X_0$.

In this paper we only consider the contribution of the diagram class shown in Fig.~\ref{fig::diags1}
to $F_{\rm box1}$ and $F_{\rm box2}$.


\section{Asymptotic expansion}
\label{sec:asyexp}

For the computation of the two-loop integrals we follow two approaches, which
we describe in the following.  For this purpose it is convenient to distinguish
the mass of the final-state Higgs bosons ($m_H^{\rm ext}$) from that of the Higgs
boson which propagates in the loops ($m_H^{\rm int}$).
This means that for the process
$gg\to HH$ we have the following dimensionful quantities: the Mandelstam
variables $s$ and $t$, and the masses $m_t$, $m_H^{\rm int}$ and
$m_H^{\rm ext}$.  In our two approaches we assume the following hierarchies:
\begin{itemize}
\item[(A)] $s,t \gg m_t^2 \gg (m_H^{\rm int})^2, (m_H^{\rm ext})^2$,
\item[(B)] $s,t \gg m_t^2 \approx (m_H^{\rm int})^2 \gg (m_H^{\rm ext})^2$.
\end{itemize}

In approach~(A) we treat the inequality $m_t^2 \gg (m_H^{\rm int})^2$ at the
level of the integrand by applying the hard-mass expansion procedure as implemented in
the program {\tt exp}~\cite{Harlander:1998cmq,Seidensticker:1999bb}.  For each
Feynman diagram this leads to two subgraphs: the two-loop diagram itself and
the one-loop diagram which contains all top quark lines. In the latter case the
co-subgraph consists only of the Higgs boson propagator.

The two-loop subgraph is Taylor-expanded in $m_H^{\rm int}$ whereas the
one-loop subgraph is expanded in the loop momentum of the co-subgraph, which
is a one-loop vacuum integral with mass scale $m_H^{\rm int}$.  In addition,
each subgraph is then expanded in $m_H^{\rm ext}$, which is performed at the level of
scalar integrals with the help of {\tt LiteRed}~\cite{Lee:2012cn,Lee:2013mka}.

At this point one has to deal with one- and two-loop four-point integrals
which only depend on the variables $s$, $t$ and $m_t$. These integrals belong to the
same set of topologies used in the calculation of the QCD corrections presented in
Refs.~\cite{Davies:2018ood,Davies:2018qvx}; we are able to re-use those results here.

Approach~(B) has the advantage that all expansions for the hierarchy 
$m_t^2 \approx (m_H^{\rm int})^2 \gg (m_H^{\rm ext})^2$ are simple Taylor
expansions; no expansion by {\tt exp} is necessary.
To implement the approximation $m_t^2 \approx (m_H^{\rm int})^2$, we write
the Higgs boson propagator in the form
\begin{eqnarray}
  i [D_h(p)]^{-1} &=& (m_H^{\rm int})^2 - p^2 \nonumber\\
  &=& m_t^2(1 - \delta^\prime) - p^2\,, 
\end{eqnarray}
where $\delta^\prime = 1-(m_H^{\rm int})^2/m_t^2$, and expand $D_h(p)$ in the limit
$\delta^\prime \to 0$ at the level of the integrand.
The expansion in $m_H^{\rm ext}$ is then performed in the same way as for
approach~(A), described above.
The remaining integrals are two-loop four-point integrals with massless legs,
where all internal propagators have the mass $m_t$; this is a different set of
integral topologies to those of the QCD corrections and approach~(A).

In the final result, it is advantageous to introduce $\delta= 1 - m_H^{\rm int}/m_t$.
By making the replacement
\begin{eqnarray}
  \delta^\prime = \delta ( 1 + m_H^{\rm int}/m_t ) = \delta ( 2 - \delta )
  \,,
  \label{eq::del_vs_delp}
\end{eqnarray}
we obtain an expansion in $\delta$ which often has better convergence
properties than the expansion in $\delta^\prime$ (see also discussion at the
end of Section~2 in Ref.~\cite{Fael:2022frj}).


\section{\label{sec::amplitude}$gg\to HH$ amplitude and form factors}

In this section we provide some details regarding how the two expansion approaches
discussed in Section~\ref{sec:asyexp} are implemented.
We generate the amplitude with {\tt qgraf}~\cite{Nogueira:1991ex} and process
the output with {\tt q2e} and {\tt
  exp}~\cite{Harlander:1998cmq,Seidensticker:1999bb} in order to generate {\tt
  FORM}~\cite{Ruijl:2017dtg} code for the amplitudes.
This yields 6 one-loop diagrams and 60 two-loop diagrams.

As mentioned above, in
approach (A) {\tt exp} identifies a one- and a two-loop sub-graph for each of
the two-loop diagrams. The corresponding four-point integrals are expanded
in $m_H^{\rm ext}$ using {\tt LiteRed}~\cite{Lee:2012cn,Lee:2013mka} and then
integration-by-parts (IBP) reduced to a set of master integrals using \texttt{FIRE}~\cite{Smirnov:2014hma}.
These master integrals, which depend on $s,t$ and $m_t$, are well-studied in the
literature and the results of Refs.~\cite{Davies:2018ood,Davies:2018qvx} can be
re-used here.

In approach~(B), {\tt exp} does not perform any expansion but simply maps each diagram
to a predefined integral family with massive final-state Higgs bosons and an internal
Higgs boson propagator with mass $m_H^{\rm int}$.
These integrals are expanded in $\delta'$ at the level of the integrand by \texttt{FORM},
and the resulting scalar integrals are expanded in $m_H^{\rm ext}$ by {\tt LiteRed} and
IBP reduced using \texttt{FIRE}. The number of master integrals is minimized using
the \texttt{FIRE} command \texttt{FindRules}, which equates identical integrals which
belong to different integral families; this procedure yields a basis of 167 master integrals.
We also apply \texttt{FindRules} to the entire list of unreduced integrals, as discussed in
Ref.~\cite{Davies:2018qvx}. Applying the IBP reduction tables to the equalities found here
yields an additional 27 non-trivial relations between master integrals, thus we finally
obtain a basis of 140 two-loop master integrals. We additionally perform the IBP reduction
of a set of test integrals using \texttt{Kira}~\cite{Maierhofer:2017gsa,Klappert:2020nbg};
here we also find a basis of 140 two-loop master integrals after minimizing between the
different families.

\begin{figure}[t]
  \centering
  \begin{tabular}{cc}
    \includegraphics[width=0.3\textwidth]{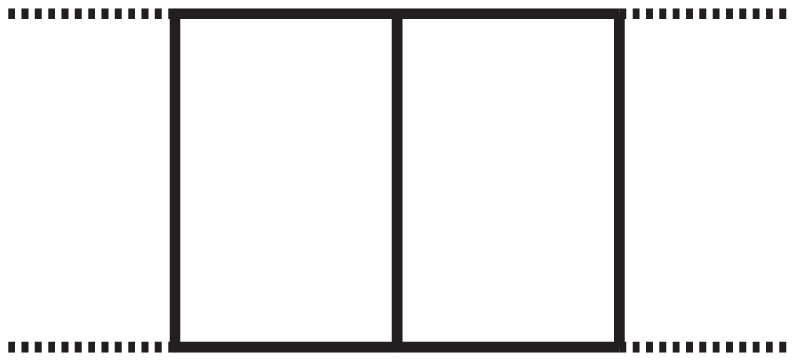}
    &
    \includegraphics[width=0.3\textwidth]{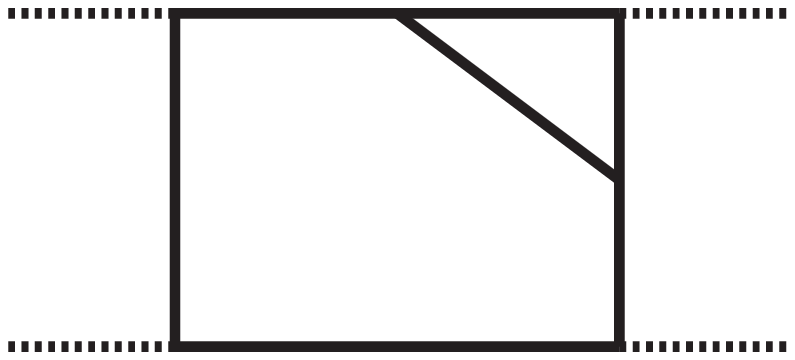}
  \end{tabular}
  \caption{\label{fig::2l-topos} The two-loop integral families which appear in expansion
  approach (B). The solid lines have mass $m_t$ and the dotted external legs are massless.
  Variations of these families with permutations of the external legs also appear.}
\end{figure}

These master integrals are four-point integrals with massless external legs, and all
propagators have the mass $m_t$. Up to permutations of the external
momenta, they belong to one of two integral families, shown in Fig.~\ref{fig::2l-topos}.
The computation of these master integrals in the limit $s,t \gg m_t^2$ is described in
Section~\ref{sec::full-mt-MI}.

The amplitudes for the two form factors are linear combinations of the master integrals,
and we expand their coefficients to order $(m_H^{\rm ext})^4$ and $(\delta')^3$. This
expansion depth requires the IBP reduction of around 350,000 scalar integrals. We also
pre-expand the coefficients in $m_t$ and $\epsilon$, and the final expansions of the
form factors are obtained after inserting the $m_t$- and $\epsilon$-expanded master integrals.

The freedom in the choice of basis for the master integrals can lead to some
undesirable properties; the first is that the denominators of the coefficients of the
master integrals
in the reduction rules do not factorize in the dimensional regulator $\epsilon$
and the kinematic invariants and masses, $s$, $t$ and $m_t$. The second is that
the coefficients contain poles in $\epsilon$, which imply that the master integrals
need to be computed to higher orders in $\epsilon$, to produce the finite contribution
of the amplitude. The first point complicates the reduction and subsequent expansions
of the amplitude, leading to poor computational performance. The second leads to
unnecessarily difficult master integral computations involving functions and
constants of higher transcendental weight which, ultimately, will cancel in the
physical amplitude.
We use an improved version of the program \texttt{ImproveMasters.m}~\cite{Smirnov:2020quc} 
to find, for each family, a basis of master integrals for which both of these
issues are avoided.


\section{\label{sec::full-mt-MI}Fully massive two-loop box master integrals}

The main purpose of this section is to provide details on the computation
of the master integrals, which is based on differential
equations~\cite{Kotikov:1990kg,Bern:1993kr,Remiddi:1997ny,Gehrmann:1999as}.
The technically most challenging part is the computation of the boundary
conditions which is described in Subsection~\ref{sub::BCs}.


\subsection{Differential equations}

The master integrals have a non-trivial dependence on two scaleless parameters
$\hat{t}=t/s$ and $\hat{m}_t^2=m_t^2/s$.
We use \texttt{LiteRed} in combination with \texttt{FIRE} to derive linear 
systems of coupled differential equations with respect to each of the variables.
In principle one can try to solve these sets of differential equations analytically,
however the results are not expressible in terms of iterated integrals but rather
involve more complicated structures like elliptic integrals.
To obtain, nevertheless, precise and easy-to-evaluate results we follow the 
ideas of Ref.~\cite{Davies:2018ood,Davies:2018qvx} and evaluate the master integrals 
analytically in the high-energy expansion, i.e., for $m_t^2 \ll s,t$.

To construct the asymptotic expansion we insert a power-log ansatz for each
master integral
\begin{eqnarray}
  I_{n} &=& \sum\limits_{i=-2}^{i_{max}} \sum\limits_{j=-4}^{j_{max}} \sum\limits_{k=0}^{i+4} c_{ijk}^{(n)}
  \epsilon^i \hat{m}_t^{2j} \log^k(\hat{m}_t^2) \,,
            \label{eq::ansatz}
\end{eqnarray}
into the differential equation in the variable $\hat{m}_t^2$ and re-expand in
$\epsilon$ and $\hat{m_t}$.  Since there are no spurious poles in $\epsilon$
either in the physical amplitudes or the differential equations we can choose
$i_{max}=0$ for all master integrals. We have produced the expansion up to
$j_{max}=120$ for each of the master integrals, however the amplitudes contain
spurious negative powers in $m_t$ in the coefficients of the master integrals
and additionally, one factor of $m_t^2$ is moved to the prefactor $\alpha_t$.
Thus the final expansion depth of the form factors for approaches~(A) and~(B)
are $\hat{m}_t^{112}$ and $\hat{m}_t^{114}$, respectively.

After inserting the ansatz given in Eq.~(\ref{eq::ansatz}) for each master
integral into the differential equation we can compare the coefficients of
$\epsilon$, $\hat{m}_t$ and $\log(\hat{m}_t^2)$ between the right- and
left-hand side of the differential equation to obtain a system of linear
equations for the expansion coefficients $c_{ijk}^{(n)}$.  We use
\texttt{Kira} together with
\texttt{FireFly}~\cite{Klappert:2019emp,Klappert:2020aqs} to solve this system
of equations in terms of a minimal set of boundary conditions, making
sure to favour coefficients which belong to simpler master integrals and low
$\epsilon$ expansion depth in the reduction.  The main challenge is then to
compute the remaining undetermined set of boundary conditions, which still
depend on the second kinematic variable $\hat{t}$. We note that the
set of boundary conditions required is independent of the value of $j_{max}$,
thus the final expansion depth of the master integrals is limited only by
the ability of \texttt{Kira} and \texttt{FireFly} to solve the large system
of equations generated by high values of $j_{max}$. Deeper expansions
than we have presented here are certainly possible, if required.

For the calculation of the $\hat{t}$ dependence of the boundary conditions in
the limit $\hat{m}_t \to 0$ we use the methods developed in
Refs.~\cite{Davies:2018ood,Davies:2018qvx,Mishima:2018olh}.  In particular, we
use the method of expansion-by-regions~\cite{Beneke:1997zp,Smirnov:2002pj} to
obtain integral representations for the required boundary coefficients.  These
are subsequently solved with the help of Mellin-Barnes integrals, either by
analytically solving summations over residues or by high-precision numerical
evaluations together with the \texttt{PSLQ} algorithm \cite{PSLQ}.  In the
following section we will describe in detail how to obtain the integral
representations of the asymptotic expansion in general and show details of the
calculation of a few examples explicitly.

We have performed numerical cross checks for all 140 master integrals with the
help of {\tt FIESTA}~\cite{Smirnov:2021rhf}.  Using Euclidean kinematics,
where both $s$ and $t$ are negative, we typically obtain six digits of agreement
for the real-valued master integrals belonging to the integral families in
Fig.~\ref{fig::2l-topos}. For these checks we use $m_t=173$~GeV and set $s=t/2$ with
$\sqrt{s}\ge 1200$~GeV. In the physical region, where $s>0$ and $t\le-s/2<0$,
we find agreement for all 140 master integrals within the numerical
uncertainty of {\tt FIESTA} which provides between two and six significant digits.
The lowest precision is obtained for the seven-line master integrals with dots,
which are numerically very challenging.  We have also performed consistency
checks by inserting our analytic high-energy expansions into the system of
$t$-differential equations and found that they are satisfied, order-by-order
in $m_t$.


\subsection{\label{sub::BCs}Boundary conditions: Mellin-Barnes approach}

In this subsection, we demonstrate the Mellin-Barnes (MB) approach for the
calculation of the boundary conditions for the $\hat{m}_t$-differential
equations for the master integrals.  We only consider the subset of master
integrals for which the Euclidean region is defined by $S,T>0$ and $U<0$,
where $S=-s, T=-t$ and $U=-u$.  The
remaining master integrals can then be found by crossing relations.
The analytic continuation to the physical region is done at the end of the calculation.


\subsubsection{Basics of Mellin-Barnes representations and template integrals}

We start with a short review of the basics of MB representations and the usage
of so-called ``template integrals'' in the asymptotic $m_t$ expansion.\footnote{For a more
  detailed discussion of the MB method, we refer to~\cite{Mishima:2018olh}.}
For a two-loop master integral with $n$ lines we employ the following $\alpha$ representation,
\begin{eqnarray}
\label{eq:alpharep}
\mathcal{I}_n (S,T,U,m_t^2) &=&
\int  \prod_{j=1}^2 \,  \rd l_j   \, \f{1}{D_1^{1+\delta_1} \cdots D_n^{1+\delta_n}} = \int_0^\infty \lb \prod_{i=1}^n \rd \alpha_i \, \f{\alpha_i^{\delta_i}}{\Gamma(1+\delta_i)} \rb \mathcal{U}^{-d/2} \, e^{\cal -F/U} , \nonumber \\
\end{eqnarray} 
where $\mathcal{U}$ and $\mathcal{F}$ are Symanzik polynomials,
$\delta_i$ are additional regulators associated with the 
denominators $D_i$, and the integration measure is chosen as 
\begin{eqnarray}
\label{eq:measure}
\int \rd l_j &:=& \f{ 1 }{\ri \, \pi^{d/2}} \, \int  \rd^d l_j \; \quad \mbox{with } \; d = 4 - 2\, \epsilon \,.
\end{eqnarray}
For later convenience, we further adopt the notation for the $\alpha$-parameter measure as
\begin{eqnarray}
\int \rd^n \alpha^\delta &:=& \int_0^\infty \prod_{i=1}^n \rd \alpha_i \, \f{\alpha_i^{\delta_i}}{\Gamma(1+\delta_i)} \;.
\end{eqnarray}

We realize the asymptotic expansion in the high-energy region with the help
of version~2.1 of the program {\tt asy}~\cite{Pak:2010pt}.
Using as input
\begin{eqnarray}
\label{eq:scaling}
m_t^2 \sim \chi\,,\; S \sim 1\,,\; T \sim 1\,,\; U \sim 1 \, \quad \mbox{with }\; \chi \ll 1\,,
\end{eqnarray}
{\tt asy} provides the possible scalings of the $\alpha$ parameters in the asymptotic expansion;
it provides a list of replacements $\alpha_i \to \chi^{n_i} \alpha_i$ which describe the
different regions contributing to the asymptotic expansion.

In the ``hard region'' we have $m_t^2 \sim \chi$, while all $\alpha$ parameters scale
as 1. Therefore, it corresponds to a simple Taylor expansion in $m_t$ which can
be realized via
\begin{eqnarray}
\label{eq:hard}
  \mathcal{I}_n^{(\mathrm{hard})} 
  &=& 
      \sum_{k=0}^{\infty} \f{\big(\chi \, m_t^2\big)^k}{k!} \, \f{\partial^k}{\partial (m_t^2)^k}\mathcal{I}_n\bigg|_{m_t^2=0} \,.
\end{eqnarray}
The integrals on the r.h.s.~can be reduced to known massless master integrals
(see, e.g., Ref.~\cite{Smirnov:1999gc,Tausk:1999vh}) using IBP methods.

For the ``soft regions'', i.e.~the regions in which at least one of the $\alpha$ parameters 
scales $\sim \chi$, we can expand the $\alpha$ representation of Eq.~\eqref{eq:alpharep} according
to the region's $\alpha$-parameter scaling as\footnote{Note that in general other scalings
are possible for which Eq.~\eqref{eq:AsyHigh} is not valid, however in the problem at hand we only
encounter regions in which the $\alpha$ parameters scale as $\chi$ or as 1.}
\begin{eqnarray}
\label{eq:AsyHigh}
\mathcal{I}_n^{(\mathrm{soft})} &=& 
\sum_{r=1}^R \sum\limits_{k=0}^\infty \int \rd^n \alpha^\delta \,
\f{\big(\chi\big)^k}{k!} \, \left[\f{\partial^k}{\partial \chi^k} \left\{\mathcal{U}_{(r)}^{-d/2} \exp\left(-\mathcal{F}_{(r)}/\mathcal{U}_{(r)}\right)\right\}\right]_{\chi=0}
\,.
\end{eqnarray}
$\mathcal{U}_{(r)}$ and $\mathcal{F}_{(r)}$ are the Symanzik polynomials
where $\chi$ has been introduced by applying 
the scaling of region $r$.
Note that contrary to the hard region, which always starts at $\mathcal{O}(m_t^0)$,
the soft regions can have different leading powers.

Taking the derivatives w.r.t.~$\chi$ in Eq.~(\ref{eq:AsyHigh})
essentially produces the content of the curly brackets multiplied by
polynomials in $\alpha_i$, dimensionful quantities, the dimension $d$,
and negative powers of ${\cal U}_{(r)}$.
This allows us to define ``shift operators'' $\hat{\calS}^k_r$
which reproduce the $k^{\rm th}$ derivative in the region $r$ without computing
the derivative explicitly. Schematically, these shift operators can be written as
\begin{eqnarray}
  \label{eq:shiftOP1}
  \hat{\calS}^k_r \Big( \{ v_j\}, \{\alpha_i\} \Big)  &=& 
  \sum_{\sigma} \left[ {\{v_j\} \, \mbox{monomial} } \right]_\sigma
  \times \frac{ \left[ \{\alpha_i\} \, \mbox{polynomial} \right]_\sigma}{(\mathcal{U}_r)^{\rho_\sigma}}
 \;,
\end{eqnarray}
where $\sigma$ runs over the various combinations of at most $k^{\rm th}$ order monomials constructed
from $v_j \in \{ m_t^2, d, S,T,U \}$, $\rho_\sigma \geq 0$ is an integer, and we have introduced
the notation:
\begin{align}
  \mathcal{U}_{r} &= \left. \mathcal{U}_{(r)} \right|_{\mbox{\small coefficient of the leading term in }\chi}, \nonumber\\
  \mathcal{F}_{r} &= \left. \mathcal{F}_{(r)} \right|_{\mbox{\small coefficient of the leading term in }\chi}.
\end{align}
The $\chi$-expansion of a region $r$ can now be interpreted, not in terms of derivatives,
but as the shifting of the indices of the single template integral of the region,
$\calT_{r}$. This template integral represents the leading integral in the region's $\chi$-expansion
and is given by
\begin{eqnarray}
  \label{eq:template}
  \calT_{r} (\{ \delta_i \}, \epsilon) 
  &:=& \int \rd^n \alpha^\delta \,
       \mathcal{U}_r^{-d/2} \, e^{-\mathcal{F}_r/\mathcal{U}_r} \;.
\end{eqnarray}
We provide {\tt Mathematica} expressions for all template integrals in
the ancillary files~\cite{progdata}.
The action of one possible term of the shift operators on the template integrals is given by:
\begin{eqnarray}
\label{eq:ShiftRule}
\hat{\calS}^k_r \Big( \{v_j\}, \{\alpha_i\} \Big)    \circ \calT_{r} \big(\{ \delta_i \}, \epsilon \big) 
&\supset& \{v_j\}\text{ monomial}\times \frac{\prod_{i=1}^n \alpha_i^{\beta_i}}{(\mathcal U_r)^\rho} \, \calT_{r} \big(\{ \delta_i \}, \epsilon \big) \nonumber \\[2mm]
&=&
 \{v_j\}\text{ monomial}\times
 \bigg( \prod_{i=1}^{n} \calP_{1+\delta_i}^{\beta_{i}} \bigg)
 \, \calT_{r} \Big(\{ \delta_i +\beta_{i} \}, \epsilon-\rho \Big) \;, \nonumber \\
\end{eqnarray}
where $\beta_i\ge0$ and $\rho\ge0$ are integers and 
$ \calP_{1+\delta_i}^{\beta_i} = \Gamma(1+\delta_i+\beta_i) /
\Gamma(1+\delta_i)$ is the Pochhammer function.

In this way, the higher-order $\chi$-expansion terms for the master integrals
without numerators\footnote{The shifting rule for master integrals with dotted
  propagators can be obtained directly from Eq.~\eqref{eq:ShiftRule}
  by changing the $\delta$ indices.}  can be obtained from a single template
integral per region.  The full expansion of a master integral in the soft
regions can therefore be written as
\begin{eqnarray}
\label{eq:MBstandard}
\mathcal{I}_n^{(\mathrm{soft})} &=& 
\sum_{r=1}^R
\Bigg[ 1 + \sum_{k=1}^{\infty}\, \chi^k    
                                    \hat{\mathcal{S}}^k_r \Big( \{v_j\}, \{\alpha_i\} \Big) \Bigg]  
\circ \calT_{r} \big(\{ \delta_i \}, \epsilon \big)  \;.
\end{eqnarray}

The MB representation of the template integrals can be obtained by means of
direct integration over the $\alpha$ parameters and the application of Mellin-Barnes 
representations,
\begin{eqnarray}
(x+y)^\lambda &=&  \int_{- \ri \infty}^{+ \ri \infty} \f{\rd z}{2 \pi \ri} \, \f{\Gamma(-\lambda+z) \, \Gamma(-z)}{\Gamma(-\lambda)} \, x^{z} \, y^{\lambda-z} \,,
\end{eqnarray}
where the integration path has to be chosen in such a way as to separate the poles 
of the $\Gamma(\dots+z)$ and $\Gamma(\dots-z)$ factors.
Note that the individual template integrals contain spurious
poles in the regulators $\delta_i$, which cancel in the sum
of all soft regions. 


\subsubsection{Mellin-Barnes representations for master integrals with numerators}

In the following we introduce a parametric method to directly obtain the MB
representations for the boundary conditions of master integrals with
numerators.  Another method would be to reduce master integrals with
numerators to a basis of master integrals with only dots via IBP reductions.
However in such a basis, deeper expansions in $\epsilon$ and $m_t$ are often
required due to the presence of spurious poles in the IBP relations.\footnote{We
notice a similar approach in \cite{Agarwal:2020dye} for numerical
evaluations of quasi-finite master integrals.}

The numerators in the $\alpha$ representation can be introduced on the same
footing as propagator denominators \cite{Borowka:2015mxa,Lee:2013mka} via
\begin{eqnarray}
\label{eq:Prop2Alpha}
\f{1}{(D_i)^\lambda} &=& \begin{cases}
\displaystyle \; \f{1}{\Gamma(\lambda)} \int_0^\infty \rd \alpha \,
\alpha^{\lambda-1} \, e^{-D_i \, \alpha} \;  & \text{for}\; \lambda >0 \,,  \\[6mm]
\displaystyle \; (-1)^{|\lambda|} \, \f{\partial^{|\lambda|}}{\partial \alpha^{|\lambda|}} \, e^{-D_i \, \alpha}\, \Big|_{\alpha=0} \;  & \text{for}\;   \lambda <0 \;.
\end{cases} 
\end{eqnarray}
The $\alpha$ representation of the $n$-line master integral with $m$ additional numerators
can then be obtained as
\begin{eqnarray}
\label{eq:alphaNum}
 \calI_{n,m}  &: =& \int  \prod_{j=1}^{2}  \rd l_j   \, \f{N_1^{\lambda_1} \cdots N_m^{\lambda_m} }{D_1^{1+\delta_1} \cdots D_n^{1+\delta_n}} 
\nonumber\\
 &=&
 \int_0^\infty \rd^n \alpha^\delta \,
\lbe \lb \prod_{t=1}^{m}  \, (-1)^{|\lambda_t|} \f{\partial^{|\lambda_t|}}{\partial \alpha_{n+t}^{|\lambda_t|}} \rb \,  \mathcal{\tilde U}^{-d/2} \, e^{\cal -{\tilde F}/{\tilde U}} \rbe_{\alpha_{n+1}=\cdots=\alpha_{n+m}=0}  \nonumber \\[2mm]
&=&   \int_0^\infty\rd^n \alpha^\delta  \;
  \mathcal{U}^{-d/2} \, e^{\cal -F/U}  \, \hat{\calO}^m \Big( \{v_j\},\{\alpha_i\} \Big) \;,
    \label{eq::Inm}
\end{eqnarray}
where in our case we have $m=1,2$. In the second line,
$\mathcal{\tilde U}$ and $\mathcal{\tilde F}$ are Symanzik polynomials in
terms of $(n+m)$ $\alpha$ parameters, while in the last line $\mathcal{U}$
and $\mathcal{F}$ are those of Eq.~\eqref{eq:alpharep} in terms of only $n$
$\alpha$ parameters. The function $\hat{\mathcal{O}}^m$ comes from the
derivatives in the second line; it has a similar form as the shift operators of
Eq.~\eqref{eq:shiftOP1}. Note that in Eq.~(\ref{eq::Inm}) no expansion in
$\chi$ has been performed.

At this stage, having derived the $n$-dimensional $\alpha$ representation, we
are ready to apply all the techniques developed for the $n$-line master
integrals to Eq.~\eqref{eq:alphaNum}.  By performing the asymptotic expansions
as described in Eq.~\eqref{eq:scaling}, the resulting hard-region integral
$\calI_{n,m}^{\mathrm{(hard)}}$ can be solved in the same way as
Eq.~\eqref{eq:hard}, and the integrals in the soft regions can be expressed as
\begin{eqnarray}
\label{eq:MBnum}
\calI_{n,m}^{(\mathrm{soft})} &=& \sum_{r=1}^R
 \int_0^\infty\rd^n \alpha^\delta  \,
  \mathcal{U}_r^{-d/2} \, e^{-\mathcal{F}_r/\mathcal{U}_r} \, \Bigg[ \sum_{k=0}^{\infty} \,  \chi^k \,\hat{\calS}_{r}^{k+m} \Big( \{ v_j \},\{\alpha_i\} \Big) \Bigg] \nonumber \\
  &=&
   \sum_{r=1}^R \, \Bigg[ \sum_{k=0}^{\infty} \,  \chi^k \,\hat{\calS}_{r}^{k+m} \Big( \{v_j\},\{\alpha_i\} \Big)  \Bigg]
   \circ  \calT_{r} \big(\{ \delta_i \}, \epsilon \big) \;,
\end{eqnarray}
where the action of the expanded shift operator $\hat{\calS}_{r}^{k+m}$
follows the same rule as in Eq.~\eqref{eq:ShiftRule}, and the template integrals
$\calT_{r}$ are the same $n$-line integrals defined in Eq.~\eqref{eq:template}.
We emphasize that the shifts from operators $\hat{\calS}_{r}^{m}$ yield the leading-order
terms in the asymptotic $m_t$ expansions of these master integrals with numerators.

Eq.~\eqref{eq:MBnum} provides an algorithmic way to obtain the MB
representations of master integrals with arbitrary numerators. Compared with
using IBP reduction to change to a basis of master integrals without
numerators, our method has the advantage of avoiding spurious higher-order poles in
$\epsilon$ and $m_t$.
Hence, one obtains a much more compact expression in terms of MB
integrals, and the cancellation of $\delta_i$-poles among
different regions can be obtained more easily.\footnote{Note that the complexity of Eq.~\eqref{eq:MBnum} at
$\ord(\chi^k)$ is similar to the $\ord(\chi^{m+k})$ expansions in
Eq.~\eqref{eq:MBstandard}.}


\subsubsection{Solving Mellin-Barnes integrals}

In order to solve the MB representations derived in Eqs.~\eqref{eq:MBstandard} and
\eqref{eq:MBnum}, the first step is to fix the integration contour and perform
analytic continuation and series expansions in the $\delta_i$ and $\epsilon$
regulators accordingly.\footnote{For details on the analytic continuation of multiple
regulators, we refer to \cite{Mishima:2018olh}.}
This step can be performed with the help of the
package \texttt{MB.m}~\cite{Czakon:2005rk}.
We now obtain a large number of multi-dimensional MB representations for
complicated integrals, which requires a systematic approach for their calculation.

In general our method aims to find infinite sums of residues
of the MB integrals, that are suitable for summation
procedures. Such residue sum representations are passed to
\texttt{EvaluateMultiSums.m} \cite{Ablinger:2010pb} and \texttt{HarmonicSums.m} \cite{HarmonicSums} which internally 
use \texttt{Sigma.m} \cite{Schneider:2007} for the analytic summation.  
This step is non-trivial, especially for multi-dimensional MB
integrals, and it involves various supplementary techniques such as adding
auxiliary scales, the $T$-expansion of MB integrals and ansatz fitting
procedures, as well as numerical evaluation and the \texttt{PSLQ} algorithm.
We will describe these methods by providing three examples in the following subsections.

In the following, we adopt the abbreviations
\begin{eqnarray}
\Gamma[x_1, \dots, x_n] \; := \; \prod_{i=1}^n \Gamma(x_i) \;, \quad \; \delta_{i_1 \dots i_n} \; := \; \sum_{m=1}^n \delta_{i_m} \;, \quad \; \alpha_{i_1 \dots i_n} \; := \; \sum_{m=1}^n \alpha_{i_m} \;,
\end{eqnarray}
and denote the Harmonic PolyLogarithms (HPLs) as $H\!\!\left(m_1,\dots,m_n,x\right)$ (see Ref.~\cite{Remiddi:1999ew} for their definition).
We use  $\log(x)$ and $H(0,x)$ interchangeably.


\subsubsection{Example 1: three-line integral}

\begin{figure}[t!]
  \centering
  \includegraphics[width=0.2\textwidth]{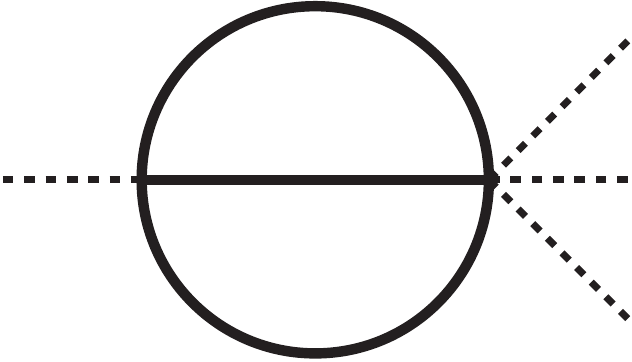}
  \caption{Three-line sunrise diagram with massless external lines. Solid
    internal lines represent massive scalar propagators. The dotted external
    lines indicate that the lines are massless.}
  \label{fig:sunrise}
\end{figure}

We start by considering the three-line massive sunrise integral with massless external lines.
The diagram is shown in Fig. \ref{fig:sunrise},
where solid and dotted lines denote massive and massless propagators, respectively.
The Symanzik polynomials are given by
\begin{eqnarray}
\mathcal U &=&\alpha _1 \alpha _2+\alpha_1 \alpha _3 +  \alpha _2 \alpha _3\;,  \nonumber \\ 
\mathcal F &=& m_t^2 \, \lb \alpha_1 + \alpha_2 + \alpha_3 \rb \, \mathcal U\;,
\end{eqnarray}
and involve only one scale, $m_t^2$; hence there is no need to perform
asymptotic expansion. We obtain a one-dimensional MB integral representation:
\begin{eqnarray}
 \mathcal I_3 &=& 
 \int \f{\rd z_1}{2 \pi \ri} \lb m_t^2\rb^{-2 \epsilon +1}  \frac{\Gamma \left[-z_1,z_1-\epsilon +2,-z_1+\epsilon -1,z_1+1,z_1+1,z_1+\epsilon \right]}{\Gamma \left[ 2-\epsilon ,2 z_1+2\right]} \,.
\end{eqnarray}
Since no expansion in $m_t$ has to be performed the regulators $\delta_i$ are not required 
and have been dropped.
We first fix the integration contour and the value of $\epsilon$ such that the
left- and right-poles of the Gamma functions are separated by a straight
line. In this case $\mbox{Re}(z_1) = -1/7$ and $\epsilon = 1$ satisfy this condition.
We then perform the analytic continuation $\epsilon \to 0$ such that we can
expand the integrand in $\epsilon$. These manipulations can be performed using
\texttt{MB.m}~\cite{Czakon:2005rk}, which yields
\begin{eqnarray}
       \label{eq:MB3line}
       \mathcal I_3 &=& (m_t^2)^{1-2\epsilon} e^{-2\epsilon\gamma_E} \Bigg(
                        -\frac{3 }{2 \epsilon ^2}
                        -\frac{9}{2 \epsilon } 
                        -\frac{21}{2}-\frac{5 \pi ^2}{12} + I^{\MB} 
                        + {\cal O}(\epsilon) \Bigg)\,,
\end{eqnarray}
where the remaining MB integral 
\begin{eqnarray}
  I^{\MB} &= & \int_{-\f{1}{7}-\ri \infty}^{-\f{1}{7}+\ri \infty} \f{\rd z_1}{2 \pi \ri} \,  \frac{\Gamma \left[-z_1-1,-z_1,z_1,z_1+1,z_1+1,z_1+2\right]}{\Gamma \left(2 z_1+2\right)} \,.
\end{eqnarray}
In order to solve the integral $I^{\MB}$ we can close the integration contour to 
the right and then sum the residues.
We obtain:
\begin{eqnarray}
I^{\MB} \!&=&  4 + \frac{\pi^2}{6}
+ 2 \sum\limits_{k=0}^\infty  \binom{2k+1}{k}^{-1} \,
\frac{\left(4 k^2+8 k+3\right) \bigl[ S_{1}(k)- S_{1}(2 k) \bigr] -4 \left(k+1\right)}{(2 k+1)(2 k + 2)(2 k+3)^2} \,, \nonumber \\
\end{eqnarray}
where $S_{i}(n) $ denote harmonic sums, i.e.,
$ S_{i}(n) = \sum_{k=1}^n \text{sign}(i)^k/{ k^{|i|}} \,.  $
As can be seen from the sum representation, the solution will be given by
inverse binomial sums at infinity which have, for example, been studied in
Ref.~\cite{Weinzierl:2004bn,Ablinger:2014bra}.  However, their associated
constants are not as well studied as those associated to the usual harmonic
sums and thus a simplification of the final result is difficult;
for this reason we proceed with a different method.
The first step is to introduce a parameter into the sum and define
\begin{eqnarray} 
I^{\MB}(\xi) \!&=&  4 + \frac{\pi^2}{6}
+ 2 \xi \, \sum\limits_{k=0}^\infty \xi^k \, \binom{2k+1}{k}^{-1} \,
\frac{\left(4 k^2+8 k+3\right) \bigl[ S_{1}(k)- S_{1}(2 k) \bigr] -4 \left(k+1\right)}{(2 k+1)(2 k + 2)(2 k+3)^2} . \nonumber \\
\label{eq:1}
\end{eqnarray}
This allows us to find a generating function of the sum in Eq.~\eqref{eq:1}
with the help of the command \texttt{ComputeGeneratingFunction} implemented in
\texttt{HarmonicSums.m}.
The result is expressed in terms of iterated integrals over the letters 
$\{ 1/x, \sqrt{4-x}\sqrt{x} \}$. 
Afterwards we rationalize the square-root valued letters with the command
\texttt{SpecialGLToH} and take the limit $\xi \to 1$ to reconstruct $I^{\MB}$ in
Eq.~\eqref{eq:MB3line}. 
The result is given by
\begin{eqnarray}
\lim_{\xi \to 1} I^{\MB}(\xi) &=&
\frac{\pi ^2}{6} + \frac{27}{2} \lb \int_0^1 \frac{1}{1+\tau _1+\tau _1^2} \, d\tau _1 \rb^2 - 9 \lb \int _0^1\int _0^{\tau _1}\frac{\tau _2}{\tau _1 \left(1+\tau _2+\tau _2^2\right)}d\tau _2d\tau _1 \rb . \nonumber \\
\end{eqnarray}
We see that the solution can be written in terms of iterated integrals with cyclotomic letters~\cite{Ablinger:2011te}.
They can be further reduced to known constants that are represented by
multiple polylogarithms
evaluated at the sixth roots of unity~\cite{Henn:2015sem} which yields, for the
$\epsilon$-finite part of the massive sunrise diagram, the result
\begin{eqnarray}
  \mathcal I_3  &=& (m_t^2)^{1-2\epsilon}  e^{-2\epsilon\gamma_E} \Bigg(
                    -\frac{3 }{2 \epsilon ^2} 
                     -\frac{9}{2 \epsilon } 
                    -\frac{21}{2}-\frac{11 \pi ^2}{12}
                    +\psi ^{(1)}\left(\tfrac{1}{3}\right) 
                        + {\cal O}(\epsilon) \Bigg) \,.
\end{eqnarray}
Here $\psi ^{(1)}\left(\frac{1}{3}\right)$ is the PolyGamma function that is
related to the Clausen function by
\begin{eqnarray}
  \mbox{Cl}_2\lb\frac{\pi }{3} \rb &=& \frac{\psi ^{(1)}\left(\frac{1}{3}\right)}{2
  \sqrt{3}}-\frac{\pi ^2}{3 \sqrt{3}}\,.
\end{eqnarray}

When reconstructing analytic expressions from numerical evaluations using the 
\texttt{PSLQ} algorithm we therefore have to to use the following basis 
of constants as well as all possible products up to transcendental weight~4:
\begin{eqnarray}
  &&\left\{1,\, \sqrt{3}, \, 
     \log (2), \, \log (3), \, \pi, \,
     \psi ^{(1)}\left(\tfrac{1}{3}\right), \,
     \zeta (3), \, 
     \right.\nonumber\\&&\left.
                          \mathrm{Im}
                          \left[\text{Li}_3\left(\tfrac{\ri}{\sqrt{3}}\right)\right],\,
                          \mathrm{Im} \left[\text{Li}_3\left(\tfrac{\ri \sqrt{3}}{4}+\tfrac{1}{4}\right)\right] ,\,
                          \text{Li}_4\left(\tfrac{1}{2}\right)
  \right\}  .
\end{eqnarray}


\subsubsection{Example 2: five-line integral}

\begin{figure}[t]
  \centering
  \includegraphics[width=0.25\textwidth]{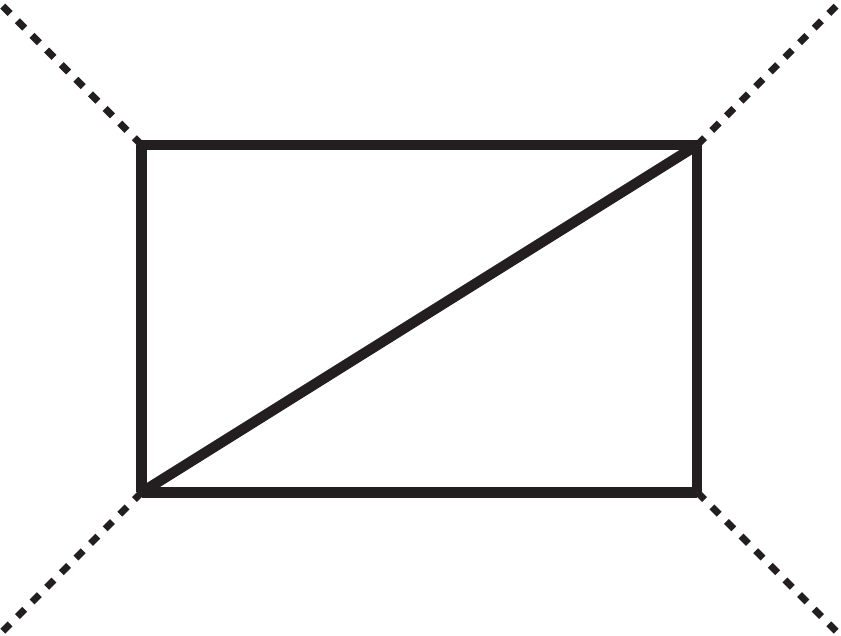}
  \caption{Five-line box diagram.}
  \label{fig:fiveline}
\end{figure}

The second example is the five-line integral shown in Fig.~\ref{fig:fiveline}.
To fix the boundary conditions for the 
$m_t$ differential equation for this integral we need to evaluate it including
$\ord(\epsilon^0,m_t^2)$;
thus asymptotic expansion beyond the leading order in the $m_t^2$ expansion is required.
The Symanzik polynomials for this master integral are given by
\begin{eqnarray}
\label{eq:UF5line}
\mathcal U &=& \left(\alpha _2+\alpha _3\right) \left(\alpha _1+\alpha _4\right)+\left(\alpha _1+\alpha _2+\alpha _3+\alpha _4\right) \alpha _5 \nonumber \\[2mm]
\mathcal F &=& 
S \,  \alpha _2 \alpha _4 \alpha _5  + T \, \alpha _1 \alpha _3 \alpha _5  + m_t^2 \, \left(\alpha _1+\alpha _2+\alpha _3+\alpha _4+\alpha _5\right) \, \mathcal U \,,
\end{eqnarray}
which involve three scales: $S,T$ and $m_t^2$. 
By applying asymptotic expansions with \texttt{asy} according to the scaling
in Eq.~\eqref{eq:scaling}, we obtain 9 regions with the following
$\alpha$-parameter scalings:
\begin{eqnarray}
\label{eq:scale5line}
(\alpha_1, \alpha_2, \alpha_3, \alpha_4, \alpha_5)  \!& \overset{\chi}{\sim} &\! \underbrace{(0, 0, 0, 0, 0)}_{(0)}, \; \underbrace{(0, 0, 0, 0, 1)}_{(1)}, \; \underbrace{(0, 0, 1, 1, 0)}_{(2)}, \; \underbrace{(0, 1, 1, 0, 0)}_{(3)}, \; \underbrace{(0, 1, 1, 0, 1)}_{(4)}, \nonumber \\[1mm]
&&  \underbrace{(1, 0, 0, 1, 0)}_{(5)}, \; \underbrace{(1, 0, 0, 1, 1)}_{(6)}, \; \underbrace{(1, 1, 0, 0, 0)}_{(7)}, \;\underbrace{(1, 1, 1, 1, 0)}_{(8)}.
\end{eqnarray}
The scaling $(0, 0, 0, 0, 0)$ corresponds to the hard region, in which only
$m_t^2 \sim \chi$ and all $\alpha$ parameters
scale as $\alpha_i \sim \chi^0$. In the remaining eight regions a subset of the $\alpha$ parameters 
scale as $\alpha_j \sim \chi$.

\paragraph{Hard region:}
For the hard region, we proceed in the same way as Eq.~\eqref{eq:hard}. The
leading term at $\ord(m_t^0)$ can be obtained by setting $m_t = 0$, which
corresponds to one of the known massless master integrals given in
Refs.~\cite{Smirnov:1999wz,Bern:2005iz}.  For the sub-leading term at
$\ord(m_t^2)$, we first perform a Taylor expansion at the integrand level, and
then perform an IBP reduction with \texttt{LiteRed} \cite{Lee:2012cn,Lee:2013mka} to
reduce again to the set of known massless master integrals to obtain the final result.

\paragraph{Soft regions:} 
For the soft regions, we apply the eight scalings from Eq.~\eqref{eq:scale5line}
to the Symanzik polynomials in Eq.~\eqref{eq:UF5line}, and expand the $\alpha$
representation to the sub-leading order in $\chi$ as described in Eq.~\eqref{eq:AsyHigh}. 
For region $(1)$ we find, for example,
\begin{eqnarray}
\label{eq:asy5line}
\mathcal{I}_5^{(\mathrm{soft}),(1)} &=& 
\int \rd^5 \alpha^\delta \,
\mathcal{U}_1^{-d/2} \, e^{-\mathcal{F}_1/{\mathcal{U}_1}} 
\Bigg[ 1  - \chi \Bigg\{
   m_t^2 \, \alpha_5 
   + \dfrac{d}{2} \, \dfrac{\alpha_{1234} \, \alpha_5}{\mathcal U_1}
  - S  \, \dfrac{\alpha _2 \, \alpha _4 \, \alpha _{1234} \, \lb \alpha _5 \rb^2}{\lb \mathcal U_1\rb^2}
  \nonumber \\ &&
  \hspace{5.5cm} - T  \, \dfrac{\alpha _1 \, \alpha _3 \, \alpha _{1234} \, \lb \alpha _5\rb^2}{\lb \mathcal U_1\rb^2}
\Bigg\}  + \ord\lb \chi^2 \rb \Bigg] \,,
\label{eq:example2_parametric}
\end{eqnarray}
with the expanded Symanzik polynomials
\begin{eqnarray}
\mathcal U_1 &=& \left(\alpha _2+\alpha _3\right) \left(\alpha _1+\alpha _4\right) \;, \nonumber \\[1mm]
\mathcal F_1 &=& S\, \alpha _2 \alpha _4 \alpha _5 + T \alpha _1 \alpha _3 \alpha _5  +
 m_t^2 \,  \big(\alpha _1+\alpha _2+\alpha _3+\alpha _4\big) \, \mathcal U_1 \;.
\end{eqnarray}
Note that $\mathcal U_1$ is the coefficient of $\chi^0$ and $\mathcal F_1$ is the
coefficient of $\chi^1$.
The eight template integrals, which correspond to the leading contributions, can
be extracted according to Eq.~\eqref{eq:template}. They are represented by, at
most, one-dimensional MB integrals.

The template integral for region $1$ is given by
\begin{eqnarray}
  \mathcal T_{1, \{\delta_1,\delta_2,\delta_3,\delta_4,\delta_5\},\epsilon} 
  &=& 
  \int \f{\rd z_1}{2 \pi \ri} \, 
  \f{\left(m_t^2\right)^{-\delta_{1234}-2 \epsilon }}{S^{\delta_5+1}}\lb \f{T}{S} \rb^{z_1}
  \frac{  \Gamma\left[ \delta_{23}+\epsilon ,\delta _{14}+\epsilon ,
      {\delta_{2} - \delta_{5} } - z_1,-z_1\right]}
      {\Gamma\left[\delta _1+1,\delta _2+1,\delta _3+1,\delta _4+1\right]} \nonumber \\[1mm]
  && \hspace{1.5cm} \times \f{\Gamma\left[{\delta_{4} - \delta_{5} } -z_1, \delta _1+z_1+1,\delta _3+z_1+1,\delta _5+z_1+1 \right]}{\Gamma\left[ \delta _{23} -\delta _5+1,\delta _1+\delta _4-\delta _5+1,\delta _5+1 \right]} \;,
\end{eqnarray}
which is obtained from Eq.~(\ref{eq:template}) through straightforward integration.
The expansion in Eq.~\eqref{eq:example2_parametric} can also be reinterpreted in terms of shift operators acting 
on this template integral:
\begin{eqnarray}
\label{eq:shift5line}
\mathcal{I}_5^{(\mathrm{soft}),(1)} 
  \!&=&\!
      \bigg[ 1 +  \chi \hspace{-4mm} \sum_{v\in\{m_t^2,d,S,T\}} \hspace{-4mm}
      \hat{\mathcal{S}}^1_1 \lb v,
      \{\alpha_i\} \rb   \bigg] 
      \circ \mathcal T_{1,\{\delta_1,\delta_2,\delta_3,\delta_4,\delta_5\},\epsilon}  \vphantom{\bigg[}\nonumber\\[-2mm]
\!&=&\!
\mathcal T_{1,\{\delta_1,\delta_2,\delta_3,\delta_4,\delta_5\},\epsilon}  + \chi \bigg[
-m_t^2 \, \mathcal{P}_{1+\delta_5}^1 \, \mathcal T_{1,\{\delta_1,\delta_2,\delta_3,\delta_4,\delta_5+1\},\epsilon} \vphantom{\bigg[}\nonumber\\[-2mm]
&&{} 
 - \f{d}{2} \, \mathcal{P}_{1+\delta_5}^1 \bigg(
\mathcal{P}_{1+\delta_1}^1  \, \mathcal T_{1,\{\delta_1+1,\delta_2,\delta_3,\delta_4,\delta_5+1\},\epsilon-1}  
+
\mathcal{P}_{1+\delta_2}^1  \, \mathcal T_{1,\{\delta_1,\delta_2+1,\delta_3,\delta_4,\delta_5+1\},\epsilon-1} \vphantom{\bigg[}\nonumber\\[-2mm]
&&{} \hspace{20mm}
+
\mathcal{P}_{1+\delta_3}^1  \, \mathcal T_{1,\{\delta_1,\delta_2,\delta_3+1,\delta_4,\delta_5+1\},\epsilon-1}
+
\mathcal{P}_{1+\delta_4}^1  \, \mathcal T_{1,\{\delta_1,\delta_2,\delta_3,\delta_4+1,\delta_5+1\},\epsilon-1}
 \bigg) \vphantom{\bigg[}\nonumber\\[-2mm]
&&{} 
 + \, S \, \mathcal{P}_{1+\delta_5}^2 \bigg(
 \mathcal{P}_{1+\delta_1}^1  \mathcal{P}_{1+\delta_2}^1  \mathcal{P}_{1+\delta_4}^1 \, \mathcal T_{1,\{\delta_1+1,\delta_2+1,\delta_3,\delta_4+1,\delta_5+2\},\epsilon-2} \vphantom{\bigg[}\nonumber\\[-2mm]
&&{} \hspace{20mm}
 +  
  \mathcal{P}_{1+\delta_2}^2  \mathcal{P}_{1+\delta_4}^1 \, \mathcal T_{1,\{\delta_1,\delta_2+2,\delta_3,\delta_4+1,\delta_5+2\},\epsilon-2} \vphantom{\bigg[}\nonumber\\[-2mm]
&&{} \hspace{20mm}
 + \,
 \mathcal{P}_{1+\delta_2}^1  \mathcal{P}_{1+\delta_3}^1  \mathcal{P}_{1+\delta_4}^1 \, \mathcal T_{1,\{\delta_1,\delta_2+1,\delta_3+1,\delta_4+1,\delta_5+2\},\epsilon-2}  \vphantom{\bigg[}\nonumber\\[-2mm]
&&{} \hspace{20mm}
 +  
  \mathcal{P}_{1+\delta_2}^1  \mathcal{P}_{1+\delta_4}^2 \, \mathcal T_{1,\{\delta_1,\delta_2+1,\delta_3,\delta_4+2,\delta_5+2\},\epsilon-2}  
 \bigg) \vphantom{\bigg[}\nonumber\\[-2mm]
&&{}
  + \, T \, \mathcal{P}_{1+\delta_5}^2 \bigg(
 \mathcal{P}_{1+\delta_1}^2  \mathcal{P}_{1+\delta_3}^1 \, \mathcal T_{1,\{\delta_1+2,\delta_2,\delta_3+1,\delta_4,\delta_5+2\},\epsilon-2} \vphantom{\bigg[}\nonumber\\[-2mm]
&&{} \hspace{20mm}
 +  
   \mathcal{P}_{1+\delta_1}^1 \mathcal{P}_{1+\delta_2}^1  \mathcal{P}_{1+\delta_3}^1 \, \mathcal T_{1,\{\delta_1+1,\delta_2+1,\delta_3+1,\delta_4,\delta_5+2\},\epsilon-2} \vphantom{\bigg[}\nonumber\\[-2mm]
&&{} \hspace{20mm}
 + \,
  \mathcal{P}_{1+\delta_1}^1  \mathcal{P}_{1+\delta_3}^2 \, \mathcal T_{1,\{\delta_1+1,\delta_2,\delta_3+2,\delta_4,\delta_5+2\},\epsilon-2}  \vphantom{\bigg[}\nonumber\\[-2mm]
&&{} \hspace{20mm}
  +
 \mathcal{P}_{1+\delta_1}^1  \mathcal{P}_{1+\delta_3}^1  \mathcal{P}_{1+\delta_4}^1 \, \mathcal T_{1,\{\delta_1+1,\delta_2,\delta_3+1,\delta_4+1,\delta_5+2\},\epsilon-2} 
 \bigg)
\bigg] \,.
\end{eqnarray}
At this point we use the MB representations derived for this region and
perform the analytic continuation and expansion of the regulators
$\delta_1, \dots, \delta_5$ and $\epsilon$, with the integration contour chosen at
$\mathrm{Re}(z_1) = -1/7$. As before, this is performed by \texttt{MB.m} and the
left- and right-poles are separated by the straight contour line.
The series expansion for the individual regions yield both $\delta_i$- and $\epsilon$-poles.
While the $\delta_i$-poles have to cancel in the sum of all the soft regions for each master 
integral, the $\epsilon$-poles cancel in the final sum of hard and soft regions for this 
diagram, since it is finite.
The resulting one-dimensional MB integrals are solved by closing the integration contours either 
to the left or right and the subsequent summation of the residue sums using \texttt{Sigma.m} and
\texttt{HarmonicSums.m} as described in the previous example.

\paragraph{Results:} Solving the MB integrals in the soft regions and combining them
with the hard region, we obtain the solution of the five-line integral of Fig.~\ref{fig:fiveline}:
\begin{eqnarray}
\mathcal{I}_5 &=&
\frac{1}{60 (S+T)} \, \bigg\{20 H\!\!\left(-1,\tfrac{T}{S}\right) \bigg[3 \left(H\!\!\left(0,\tfrac{T}{S}\right)^2+\pi ^2\right) H\!\!\left(0,\tfrac{m_t^2}{S}\right)+6 \big(H\!\!\left(0,0,-1,\tfrac{T}{S}\right) \vphantom{\bigg\{}\nonumber\\[-2mm] 
&&{}+\zeta (3)\big) - \, 2 H\!\!\left(0,\tfrac{T}{S}\right)^3-3 \left(2 H\!\!\left(0,-1,\tfrac{T}{S}\right)+\pi ^2\right) H\!\!\left(0,\tfrac{T}{S}\right)\bigg] +20 H\!\!\left(0,\tfrac{m_t^2}{S}\right) \vphantom{\bigg\{}\nonumber\\[-2mm]
&&{} \times \bigg(6 \left(H\!\!\left(0,0,-1,\tfrac{T}{S}\right)+\zeta (3)\right) -\,2 H\!\!\left(0,\tfrac{T}{S}\right)^3-3 \left(2 H\!\!\left(0,-1,\tfrac{T}{S}\right)+\pi ^2\right) H\!\!\left(0,\tfrac{T}{S}\right)\bigg) \vphantom{\bigg\{}\nonumber\\[-2mm]
&&{}+30 \left(H\!\!\left(0,\tfrac{T}{S}\right)^2+\pi ^2\right) H\!\!\left(0,\tfrac{m_t^2}{S}\right)^2 -\, 120 \zeta (3) H\!\!\left(0,\tfrac{T}{S}\right)+15 H\!\!\left(0,\tfrac{T}{S}\right)^4 \vphantom{\bigg\{}\nonumber\\[-2mm]
&&{}+30 \pi ^2 H\!\!\left(0,\tfrac{T}{S}\right)^2+60 H\!\!\left(0,-1,\tfrac{T}{S}\right) H\!\!\left(0,\tfrac{T}{S}\right)^2 +\,120 H\!\!\left(0,-1,-1,\tfrac{T}{S}\right) H\!\!\left(0,\tfrac{T}{S}\right) \vphantom{\bigg\{}\nonumber\\[-2mm]
&&{}+30 H\!\!\left(-1,\tfrac{T}{S}\right)^2 \left(H\!\!\left(0,\tfrac{T}{S}\right)^2+\pi ^2\right)-120 H\!\!\left(0,0,-1,-1,\tfrac{T}{S}\right) \vphantom{\bigg\{}\nonumber\\[-2mm]
&&{}-\,120 H\!\!\left(0,0,0,-1,\tfrac{T}{S}\right)+4 \pi ^4 \bigg\} \,+\, \frac{m_t^2}{S T} \, \bigg\{ 2 H\!\!\left(0,\tfrac{m_t^2}{S}\right)^3 + \left(2-3 H\!\!\left(0,\tfrac{T}{S}\right)\right) \vphantom{\bigg\{}\nonumber\\[-2mm]
&&{} \times H\!\!\left(0,\tfrac{m_t^2}{S}\right)^2 +\, \left(-2 H\!\!\left(0,\tfrac{T}{S}\right)-3 \pi ^2-8\right) H\!\!\left(0,\tfrac{m_t^2}{S}\right)+H\!\!\left(0,\tfrac{T}{S}\right)^3 \vphantom{\bigg\{}\nonumber\\[-2mm]
&&{}+\left(6 H\!\!\left(0,-1,\tfrac{T}{S}\right)+3 \pi ^2+4\right) H\!\!\left(0,\tfrac{T}{S}\right) +\, H\!\!\left(-1,\tfrac{T}{S}\right) \left(-3 H\!\!\left(0,\tfrac{T}{S}\right)^2-3 \pi ^2\right) \vphantom{\bigg\{}\nonumber\\[-2mm]
&&{}-6 H\!\!\left(0,0,-1,\tfrac{T}{S}\right)-14 \zeta (3)-\pi ^2 \bigg\} \,+\, \ord(\epsilon, m_t^4)\,, \vphantom{\bigg\{}
\end{eqnarray}
which is free from $\delta_i$- and $\epsilon$-poles.


\subsubsection{Example 3: seven-line integral with two numerators}

\begin{figure}[hbt!]
  \centering
  \includegraphics[width=0.25\textwidth]{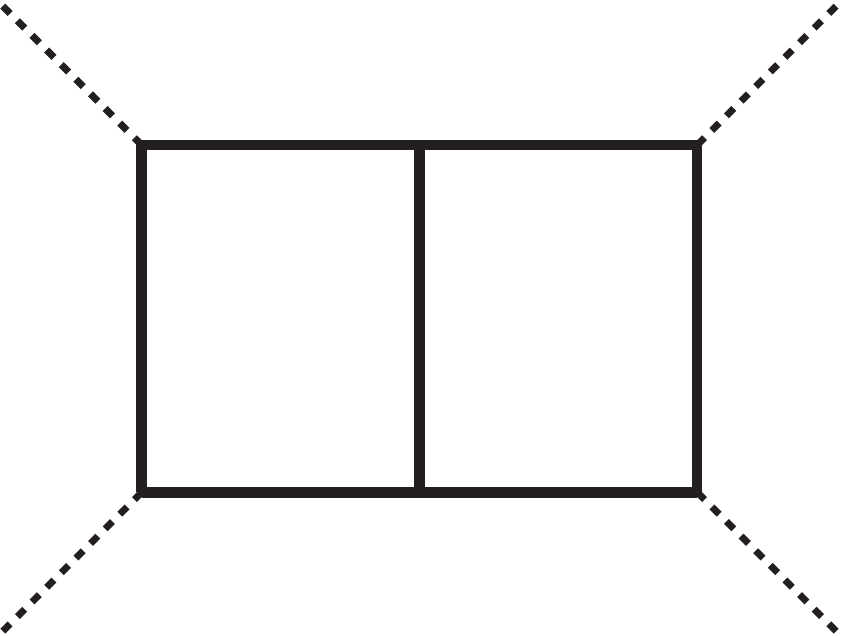}
  \caption{Seven-line double box diagram.}
  \label{fig:DBox}
\end{figure}

As a final example we consider the seven-line double box integral (see
Fig.~\ref{fig:DBox}) with two additional numerators, which needs to be evaluated to
$\ord(\epsilon^0,m_t^0)$ for the boundary conditions.\footnote{It corresponds
  to the integral $\tt G[4, \{1,1,1,1,1,1,1,-1,-1\}]$ in the ancillary file
  to this paper~\cite{progdata}.}

\paragraph{Alpha representation:} We first derive the $\alpha$ representation of
this ``7+2''-line integral by the method presented in Eqs.~\eqref{eq:Prop2Alpha}-\eqref{eq:alphaNum} 
by treating the numerators on the same footing as the propagator denominators. The resulting Symanzik
polynomials are given by
\begin{eqnarray}
\tilde{\mathcal U} &=&
\alpha _6 \left(\alpha _3+\alpha _4+\alpha _5+\alpha _7\right)+\left(\alpha _3+\alpha _4+\alpha _5\right) \alpha _8+\alpha _7 \left(\alpha _3+\alpha _4+\alpha _5+\alpha _8\right) \vphantom{\Big(}\nonumber\\[-2mm]
&&{} \hspace{-5mm} +\left(\alpha _6+\alpha _7+\alpha _8\right) \alpha _9 
+ \, {(\alpha _1+\alpha_2)} \left(\alpha _3+\alpha _4+\alpha _5+\alpha _7+\alpha
   _9\right)
\,, \vphantom{\Big(}\nonumber\\[1mm] 
\tilde{\mathcal F} &=&
S\, \Big(
\alpha _3 \big(\alpha _5 \left(\alpha _6+\alpha _7+\alpha _8\right)+\alpha _7 \left(\alpha _6+2 \alpha _8\right)\big)+\big(\alpha _7 \alpha _8-\alpha _4 \left(\alpha _6+\alpha _7+\alpha _8\right)\big) \alpha _9 \vphantom{\Big(}\nonumber\\[-2mm]
&&{} \hspace{5mm}
+ \, \alpha _2 \big(
\alpha _7 \left(\alpha _6+2 \alpha _8\right)+\alpha _5 \left(\alpha _6+\alpha _7+2 \alpha _8\right)
+\alpha _3 \left(\alpha _5+\alpha _6+2 \alpha _8\right)
\vphantom{\Big(}\nonumber\\[-2mm]
&&{} \hspace{5mm} + \,
\alpha _4 \left(\alpha _6+2 \alpha _8-\alpha _9\right)
+ \left(\alpha _6+2 \alpha _8\right) \alpha _9\big)
+ \alpha _1 \left(\alpha _3 \alpha _5-\left(\alpha _4+\alpha _7\right) \alpha _9\right)\Big)
\vphantom{\Big(}\nonumber\\[-2mm]
&&{} \hspace{-5mm}
+ T\, \Big( \alpha _1 \big(\alpha _4 \left(\alpha _7-\alpha _8-\alpha _9\right)
- \,\alpha _8 \left(\alpha _3+\alpha _5+\alpha _7+\alpha _9\right)\big)
-\big(\alpha _4 \left(\alpha _2+\alpha _6+\alpha _7\right)
\vphantom{\Big(}\nonumber\\[-2mm]
&&{} \hspace{5mm}
+\left(\alpha _4-\alpha _7\right) \alpha _8\big) \alpha _9 \Big)
\vphantom{\Big(}\nonumber\\[-2mm]
&&{} \hspace{-5mm} +\,
m_t^2 \,\Big( \alpha _1+\alpha _2+\alpha _3+\alpha _4+\alpha _5+\alpha
   _6+\alpha _7 \Big) \, \tilde{\mathcal U} \,,
\end{eqnarray}
and the seven-line Symanzik polynomials can be obtained from them:
\begin{eqnarray}
\label{eq:UF7line}
\mathcal U  \; = \; \tilde{\mathcal U}\big|_{\alpha_8=\alpha_9=0} \;,
&\quad&
\mathcal F  \; = \; \tilde{\mathcal F}\big|_{\alpha_8=\alpha_9=0} \;.
\end{eqnarray}
The $\alpha$ representation can be computed as
\begin{eqnarray}
\label{eq:alphaNum7line}
 \calI_{7,2}  &=& 
 \int_0^\infty \rd^7 \alpha^\delta \,
\lb  \f{\partial}{\partial \alpha_{8}} \f{\partial}{\partial \alpha_{9}} \,  \mathcal{\tilde U}^{-d/2} \, e^{\cal -{\tilde F}/{\tilde U}} \rb \bigg|_{\alpha_{8}=\alpha_{9}=0}  \nonumber \\[2mm]
&=&   \int_0^\infty\rd^7 \alpha^\delta  \,
  \mathcal{U}^{-d/2} \, e^{\cal -F/U} \,\bigg(  
    { \hat{\mathcal O}^2 \left( d, \{\alpha_i\} \right)
    + \hat{\mathcal O}^2 \left( S, \{\alpha_i\} \right)
    + \hat{\mathcal O}^2 \left( T, \{\alpha_i\} \right) }
    \nonumber\\[1mm]&&\mbox{}
                       + \hat{\calO}^2\! \lb d^2,\{\alpha_i\} \rb 
                       + \hat{\calO}^2\! \lb S^2,\{\alpha_i\} \rb 
                       + \hat{\calO}^2\! \lb T^2,\{\alpha_i\} \rb 
                       + \hat{\calO}^2\! \lb d\,S,\{\alpha_i\} \rb 
    \nonumber\\[1mm]&&\mbox{}
                       + \hat{\calO}^2\! \lb d\,T,\{\alpha_i\} \rb 
                       + \hat{\calO}^2\! \lb S\,T,\{\alpha_i\} \rb
\bigg),
\end{eqnarray}
Note that no expansion is $\chi$ has yet been performed.
The shift operators $\hat{\calO}^2$ originate from the numerators of the integral,
as explained in Eq.~\eqref{eq:alphaNum}, and read:
\begin{eqnarray}
\label{eq:shiftOP7line}
{  \hat{\mathcal O}^2 \left( d, \{\alpha_i\} \right) }
   &=& \frac{d}{2 \left( \mathcal U \right)^2} \; \alpha_7^2 \,, \nonumber \\[1mm]
{  \hat{\mathcal O}^2 \left( S, \{\alpha_i\} \right) }
   &=&
   \frac{S}{\left(\mathcal{U}\right)^3} \Big[
   2 \alpha _{1267} \left(\alpha _2 \alpha _{3457}+\alpha _3 \alpha _7\right) \left(\left(2 \alpha _{12}+\alpha _6\right) \alpha _{345}+\alpha _7 \left(\alpha _{56}+2 \alpha _{1234}\right)\right) \nonumber \\[1mm]
   && -\, \mathcal{U} \,\alpha _{27} \left(\left(2 \alpha _{12}+\alpha _6\right) \alpha _{345}+\alpha _7 \left(\alpha _{56}+2 \alpha _{1234}\right)\right) \nonumber \\[1mm]
   &&-\, \mathcal{U} \left(2 \alpha _{12}+\alpha _{67}\right) \left(\alpha _2 \alpha _{3457}+\alpha _3 \alpha _7\right) \Big] \nonumber \,,\\[1mm]
{   \hat{\mathcal O}^2 \left( T, \{\alpha_i\} \right) }
   &=&
   \frac{T}{\left(\mathcal U \right)^3} \Big[
   -2 \alpha _1 \alpha _{1267} \alpha _{3457} \left(\mathcal{U} + \alpha _4 \alpha _7 \right) +\alpha _1 \mathcal{U} \alpha _{1267} \alpha _{3457} \nonumber\\[1mm]
   &&+\, \left(\alpha _1-\alpha _7\right) \mathcal{U} \left( \mathcal{U}+\alpha _4 \alpha _7\right) 
   \Big] \,,
\nonumber\\[1mm]
\hat{\calO}^2 \lb d^2,\{\alpha_i\} \rb &=& \frac{d^2}{4 \lb \mathcal U \rb^2} \, {\alpha_{3457} \, \alpha_{1267} } \,, \vphantom{\frac{(T)^1}{(S)^1}}\nonumber\\[0mm]
 \hat{\calO}^2 \lb S^2,\{\alpha_i\} \rb &=& - \f{S^2}{\lb \mathcal U\rb^4} \,
 \left(\alpha_{34} \, \alpha_{126} + \alpha_{1234}\, \alpha _7\right) \left(\alpha_{45} \alpha_{126} 
 +\alpha_{1456}\, \alpha _7\right) 
  \vphantom{\frac{(T)^1}{(S)^1}}\nonumber\\[-4mm]
 && \hspace{-5mm} \times \; 
 \big(\alpha_{345}\, \left(2 \, \alpha_{12}+\alpha _6\right)+\left(2\, \alpha _{1234} + \alpha _{56} \right) \alpha _7\big) \left(\alpha _3 \alpha _7+\alpha _2 \, \alpha _{3457}\right) \,, \vphantom{\frac{(T)^1}{(S)^1}}\nonumber\\[0mm]
  \hat{\calO}^2 \lb T^2,\{\alpha_i\} \rb &=& \f{T^2}{\lb \mathcal U \rb^4} \,
  \alpha _1 \alpha _4 \alpha_{3457} \, \alpha_{1267} \big(
  \mathcal U + \alpha _1\alpha _7
  \big) 
  \big( \mathcal U + \alpha _4\alpha _7\big)
  \,, \vphantom{\frac{(T)^1}{(S)^1}}\nonumber\\[0mm]
  \hat{\calO}^2 \lb d\, S ,\{\alpha_i\} \rb &=& - \f{d\,S}{2 \lb \mathcal U \rb^3} \,
  \Big[
  \big(\left(\alpha _1 \alpha_{47}+\alpha _4 \alpha_{67}+\alpha _2 \left(\alpha _4-\alpha _6\right)\right) \, \mathcal U \, \alpha_{3457}\big) \vphantom{\frac{(T)^1}{(S)^1}}\nonumber\\[-4mm]
  && \hspace{-5mm} +\;
  \big(\alpha _2 \left(\alpha _6 \alpha_{45}+\alpha _3 \alpha_{56}+\alpha _7 \alpha_{56}\right)+\alpha _3 \left(\alpha _5 \alpha_{67}+\alpha _6 \alpha _7\right)+\alpha _1 \alpha _3 \alpha _5\big) \alpha_{1267} \alpha_{3457} \vphantom{\frac{(T)^1}{(S)^1}}\nonumber\\[-4mm]
  && \hspace{-5mm} - \;
  \big(\left(2 \alpha_{12}+\alpha _6\right) \alpha_{345}+\alpha _7 \left(2 \alpha_{1234}+\alpha_{56}\right)\big) \alpha_{1267} \left(\alpha_2 \alpha_{3457}+\alpha _3 \alpha _7\right)
  \Big]\,, \vphantom{\frac{(T)^1}{(S)^1}}\nonumber\\[0mm]
  \hat{\calO}^2 \lb d\, T ,\{\alpha_i\} \rb &=& - \f{d\,T}{2 \lb \mathcal U \rb^3} \,
  \alpha_{3457} \, \alpha_{1267} \left(2 \alpha _1 \alpha _4 \alpha _7+ \alpha_{14} \, \mathcal U\right) \,, \vphantom{\frac{(T)^1}{(S)^1}}\nonumber\\[0mm]
  \hat{\calO}^2 \lb S\, T ,\{\alpha_i\} \rb &=& \f{S\,T}{ \lb \mathcal U \rb^4} \,
  \Big(
  \alpha _1 \mathcal{U} \left(\alpha _1 \alpha _{47}+\alpha _4 \alpha _{67}+\alpha _2 \left(\alpha _4-\alpha _6\right)\right) \alpha _{3457} \left(\alpha _4 \alpha _7+\mathcal{U}\right) \vphantom{\frac{(T)^1}{(S)^1}}\nonumber\\[-4mm]
  && \hspace{-5mm} +\,
  \alpha _1 \left(\alpha _3 \alpha _5 \alpha _{12}+\alpha _7 \alpha _{23} \alpha _{56}+\alpha _6 \left(\alpha _2 \alpha _{345}+\alpha _3 \alpha _5\right)\right) \alpha _{1267} \alpha _{3457} \left(\alpha _4 \alpha _7+\mathcal{U}\right) \vphantom{\frac{(T)^1}{(S)^1}}\nonumber\\[-4mm]
  && \hspace{-5mm} +\,
  \alpha _1 \alpha _4 \alpha _7 \alpha _{1267} \left(\alpha _2 \alpha _{3457}+\alpha _3 \alpha _7\right) \left(\alpha _7 \alpha _{56}+\alpha _6 \alpha _{345}-2 \, \mathcal{U}\right) \vphantom{\frac{(T)^1}{(S)^1}}\nonumber\\[-4mm]
  && \hspace{-5mm} -\,
  \alpha _4 \alpha _{1267} \, \mathcal{U}  \left(\alpha _2 \alpha _{3457}+\alpha _3 \alpha _7\right) \left(-\alpha _7 \alpha _{56}-\alpha _6 \alpha _{345}+2\, \mathcal{U}\right)
  \Big) \,, \vphantom{\frac{(T)^1}{(S)^1}}
\end{eqnarray}
with $\,\mathcal U = \alpha_{345} \, \alpha_{126} + \alpha_{123456} \, \alpha _7$.
The absence of the remaining five possible shift operators
$\{ { \hat{\calO}^2(m_t^2), } \hat{\calO}^2(m_t^4), \hat{\calO}^2(d\,m_t^2), \hat{\calO}^2(S\,m_t^2),
\hat{\calO}^2(T\,m_t^2) \}$ is expected as the numerators are irreducible
scalar products, which are free from $m_t^2$ terms.

\paragraph{Asymptotic expansions:}
With the representation in terms of seven $\alpha$ parameters for this ``7+2''-line
integral in hand, we can again apply the asymptotic expansions for the
scaling of Eq.~\eqref{eq:scaling} to the seven-line Symanzik polynomials $\mathcal U$ and
$\mathcal F$ (see Eq.~\eqref{eq:UF7line}) as well as the shift operators
in Eq.~\eqref{eq:shiftOP7line}.
The asymptotic expansion from \texttt{asy} yields the hard region and 13
soft regions with the following scalings:
\begin{eqnarray}
\label{eq:scale7line}
(\alpha_1, \dots , \alpha_7)  
  &
    \overset{\chi}{\sim}
  & \underbrace{(0, 0, 0, 0, 0, 0, 0)}_{0}, \; \underbrace{(0, 0, 0, 0, 1, 1,
    1)}_{1}, \; \underbrace{(0, 0, 0, 1, 1, 1, 0)}_{2}, \; \underbrace{(0, 0,
    1, 0, 0, 1, 1)}_{3}, \nonumber \\[1mm]    
&& \hspace{-2.9cm} \underbrace{(0, 0, 1, 1, 1, 1, 1)}_{4}, \;  \underbrace{(0,
   1, 0, 0, 1, 0, 1)}_{5}, \; \underbrace{(0, 1, 1, 0, 0, 0, 1)}_{6}, \;
   \underbrace{(0, 1, 1, 1, 0, 0, 0)}_{7}, \; 
   \underbrace{(0, 1, 1, 1, 1,  0, 1)}_{8}, \nonumber \\[1mm] 
&&  \hspace{-2.9cm}  \underbrace{(1, 0, 0, 0, 1, 1, 0)}_{9}, \;
   \underbrace{(1, 1, 0, 0, 1, 1, 1)}_{10}, \; \underbrace{(1, 1, 1, 0, 0, 0,
   0)}_{11}, \; 
\underbrace{(1, 1, 1, 0, 0, 1, 1)}_{12}, \; \underbrace{(1, 1, 1, 1, 1, 1,
   0)}_{13}. \nonumber \\   
\end{eqnarray}
For the hard region, we proceed in the standard way, i.e. we take the
massless limit and perform IBP reductions to the known massless master integrals.
For the 13 soft regions, we expand the $\alpha$ representation
in Eq.~\eqref{eq:alphaNum7line} according to Eq.~\eqref{eq:MBnum}, 
\begin{eqnarray}
\label{eq:asy7line}
 \calI_{7,2}^{\mathrm{(soft)}}  
&=& 
\sum_{r=1}^{13}
  \int_0^\infty\rd^7 \alpha^\delta  \,
  \mathcal{U}^{-d/2}_r \, e^{-\mathcal{F}_r/\mathcal{U}_r} \,\bigg(  
{     \hat{\calS}^{{2}}_r \lb d,\{\alpha_i\} \rb 
      +  \hat{\calS}^{{2}}_r \lb S,\{\alpha_i\} \rb 
      +  \hat{\calS}^{{2}}_r \lb T,\{\alpha_i\} \rb }
    \nonumber\\[1mm]&&\mbox{}
    +  \hat{\calS}^{{2}}_r \lb d^2,\{\alpha_i\} \rb 
    +  \hat{\calS}^{{2}}_r \lb S^2,\{\alpha_i\} \rb 
    +  \hat{\calS}^{{2}}_r \lb T^2,\{\alpha_i\} \rb 
     + \hat{\calS}^{{2}}_r \lb d\,S,\{\alpha_i\} \rb 
    \nonumber \\[1mm] && \mbox{}
     + \hat{\calS}^{{2}}_r \lb d\,T,\{\alpha_i\} \rb 
     + \hat{\calS}^{{2}}_r \lb S\,T,\{\alpha_i\} \rb + \ord(\chi)
  \bigg)\,.
\end{eqnarray}
The expanded shift operator $\hat{\calS}^{{2}}_r$ is the leading term
of the operator $\hat{\calO}^{2}$ in Eq.~\eqref{eq:shiftOP7line} where the
$r^{th}$
region scales according to Eq.~\eqref{eq:scale7line}. The 13 template integrals can be
identified by $\mathcal{U}_r$ and $\mathcal{F}_r$ according to
Eq.~\eqref{eq:template}. By performing parametric integrations and Mellin
transformations, we obtain up to three-dimensional MB representations for the template
integrals.  By applying the  shift operators in Eq.~\eqref{eq:ShiftRule} to
Eq.~\eqref{eq:asy7line}, we obtain the MB representations of the soft
regions.\footnote{An explicit example of applying the shift operators is shown
  in Eq.~\eqref{eq:shift5line}.}

The next step is to perform an analytic continuation w.r.t.~the eight regulators
$\delta_1, \dots, \delta_7$ and $\epsilon$. We fix the integration contours at
$\{ \mathrm{Re}(z_1) = -1/7, \mathrm{Re}(z_2) = -1/11, \mathrm{Re}(z_3) =
-1/17 \}$ as straight lines. Then we perform the continuation with the
\texttt{MB.m} package and expand the expression to order
$\ord(\delta_i^0)$ and $\ord(\epsilon^0)$. This yields a large
number of one-, two- and three-dimensional MB integrals; 2003, 515 and 14
respectively.
In the following paragraphs, we will demonstrate our method to solve
multi-dimensional MB integrals, focussing in particular on non-trivial
examples which have a non-zero contribution from the contour-closing arc
at infinity which must be taken into account.

\paragraph{Arc and residue sums:} Here we start with a simple but non-trivial
example which appears in our calculations, which demonstrates the importance of the
arc contribution. The example is a one-dimensional scaleless MB integral with the
integrand
\begin{eqnarray}
\label{eq:exampleArc}
f (z_2) &=& 
\frac{z_2^8 \, \Gamma(-z_2)^2 \, \Gamma(z_2)^2}{(z_2+1)^3 \, (z_2+2)^3}\, ,
\end{eqnarray}
where the integration contour is fixed at $\mathrm{Re}(z_2) = -1/11$. Cauchy's
residue theorem states that
\begin{eqnarray}
\int_{-\f{1}{11}-\ri \infty}^{-\f{1}{11}+\ri \infty} \f{\rd z_2}{2 \pi \ri}
  f(z_2)  &=& {-} \sum_{k=0}^{\infty} \mathrm{Res}_{z_2=k} \left[ f(z_2) \right] - \int_{\mathrm{arc}} \f{\rd z_2}{2 \pi \ri} f(z_2) \,,
\label{eq::arc1}
\end{eqnarray}
where the ($-$) sign comes from the fact that we close the contour clockwise.
One usually assumes that the arc contribution vanishes. However, this is
not the case for Eq.~\eqref{eq:exampleArc}. Closing the integration contour 
to the right and summing the residues we obtain
\begin{eqnarray}
{-}\sum_{k=0}^{\infty} \mathrm{Res}_{z_2=k} \left[ f(z_2) \right] 
  &=&
      -\sum_{k=0}^{\infty}\frac{3 k^5 (4+3k)}{ { (1+k)^4(2+k)^4  } }
\label{eq::sum1}
\nonumber \\
&=& -18 \zeta (3)-\frac{3 \pi ^2}{2}-\frac{21 \pi ^4}{10}+240 \,.
\end{eqnarray}
On the other hand, regularizing the integrand by multiplying with $\xi^{z_2}$ and summing the 
residues we obtain
\begin{eqnarray}
  -
  \sum_{k=0}^{\infty} \mathrm{Res}_{z_2=k} \left[ \xi^{z_2} f(z_2) \right] 
  &=& 
  -\sum_{k=0}^{\infty}\xi^{k} \left( \frac{3 k^5 (4+3k)}  { { (1+k)^4(2+k)^4  } } + \frac{k^6}{(1+k)^3(2+k)^3} \log(\xi) \right)
  \nonumber \\
  &\overset{\xi \to 1}{=}& -18 \zeta (3)-\frac{3 \pi ^2}{2}-\frac{21 \pi ^4}{10}+241 ~.
\label{eq::sum2}
\end{eqnarray}
The same result can be found by precise numerical integration and employing the \texttt{PSLQ} algorithm.
The difference between the two results in Eqs.~(\ref{eq::sum1})
and~(\ref{eq::sum2}) is the missing contribution from the arc
in Eq.~(\ref{eq::arc1}):
\begin{eqnarray}
\label{eq:arc1}
 \int_{\mathrm{arc}} \f{\rd z_2}{2 \pi \ri} f(z_2) &=& -1 \,.
\end{eqnarray}

Therefore, in order to systematically take the arc contribution into account,
we always rely on numerical integration of the MB integrals accompanied by
the \texttt{PSLQ} algorithm to cross-check results obtained from the
residue summations for scaleless MB integrals.
However, the problem becomes more complicated when a non-vanishing arc
contribution like Eq.~\eqref{eq:arc1} is nested in two-dimensional MB integrals
involving the kinematic invariants $T/S$.
In the following we will introduce a method which can deal with such
situations.

\paragraph{Nested arc contribution:}
For two-dimensional MB integrals, we always first try to reduce their
dimensionality using Barnes' lemmas as implemented in
\texttt{barnesroutines.m}~\cite{barnesroutines} and other simplification tricks.
For the remaining two-dimensional MB integrals involving kinematic invariants
and a nested arc contribution, we need a more careful analysis.
Let us now consider two-dimensional MB integrals of the form
\begin{eqnarray}
\label{eq:MB2Dkin}
\int  \f{\rd z_1}{2 \pi \ri} \f{\rd z_2}{2 \pi \ri} \, \lb \f{T}{S} \rb^{z_1} \hat{\Gamma}(z_1) \, \hat{\Gamma}(z_2) \, \hat{\Gamma}(z_1,z_2) \,,
\end{eqnarray}
where $\hat{\Gamma}$ denotes the product of Gamma functions with common
integration variables. In our case we have two types of $z_1$ residues
from the Gamma functions, which are given by
\begin{eqnarray}
\label{eq:resZ1}
\begin{cases}
\mbox{type 1: }z_1 = 0, \, 1,\,  2,\, \dots \\[2mm]
\mbox{type 2: }z_1 = g(z_2), \, g(z_2)+1, \, g(z_2) +2, \, \dots
\end{cases}
\end{eqnarray}
From the type 1 residues with integer $z_1$ we obtain
\begin{eqnarray}
\label{eq:resType1}
  I_1 &=& {-}\int  \f{\rd z_2}{2 \pi \ri} \, \sum_{k_1=0}^{\infty} \mathrm{Res}_{z_1=k_1} \, \lb \f{T}{S} \rb^{z_1} \hat{\Gamma}(z_1) \, \hat{\Gamma}(z_2) \, \hat{\Gamma}(z_1,z_2) \nonumber \\[2mm]
      &=& {-}\sum_{k_1=0}^{\infty}  \lb \f{T}{S} \rb^{k_1} \,  \int  \f{\rd z_2}{2 \pi \ri} \, \hat{F}(k_1,z_2) \, \hat{\Gamma}(z_2) \, \hat{\Gamma}(k_1,z_2) \,,
\end{eqnarray}
where $\hat{F}(k_1,z_2)$ denotes the resulting residue function. 
From the type 2 residues in Eq.~\eqref{eq:resZ1}, we have
\begin{eqnarray}
\label{eq:resType2}
  I_2 &=& {-}\int  \f{\rd z_2}{2 \pi \ri} \, \sum_{k_1=0}^{\infty} \mathrm{Res}_{z_1=g(z_2)+ k_1} \, \lb \f{T}{S} \rb^{z_1} \hat{\Gamma}(z_1) \, \hat{\Gamma}(z_2) \, \hat{\Gamma}(z_1,z_2) \nonumber \\[2mm]
      &=& {-}\sum_{k_1=0}^{\infty} \,  \int  \f{\rd z_2}{2 \pi \ri} \,  \lb \f{T}{S} \rb^{g(z_2)+ k_1}  \,  \hat{F}\big(g(z_2)+ k_1,z_2\big) \, \hat{\Gamma}(z_2) \, \hat{\Gamma}\big(g(z_2)+ k_1,z_2 \big) \,.
\end{eqnarray}
We can then take the nested $z_2$ residues in Eqs.~\eqref{eq:resType1}
and~\eqref{eq:resType2}, which introduces a second infinite sum over $k_2$,
and then perform the residue summations over both $k_1$ and $k_2$ with the help of
\texttt{Sigma.m} and \texttt{EvaluateMultiSums.m}. However, this
two-dimensional $(k_1,k_2)$ residue summation will miss the arc contributions
in the first type, given in Eq.~\eqref{eq:resType1}, from scaleless
one-dimensional MB integrals in $z_2$.
The residue summation for the second type, given in Eq.~\eqref{eq:resType2},
is correct, since the kinematic scale choice $0<T/S < 1$ will suppress the
asymptotic behaviour of the integrands and ensure that the arc contributions
in Eq.~\eqref{eq:resType2} are vanishing. Instead of introducing another
regulator into the two-dimensional MB integrals, which would increase the
computational complexity significantly, we use precise numerical integration
together with the PSLQ algorithm in order to find the correct results at
fixed values of $k_2$. Clearly we can not compute the infinite sum in this
way, so we introduce the method of $T$-expansion and ansatz fitting procedures
to obtain the correct result for Eq.~\eqref{eq:MB2Dkin}.

\paragraph{Ansatz fitting and $T$-expansions:}
The basic idea of this method is to start with an ansatz for
the sum of MB integrals of the type given in Eq.~\eqref{eq:MB2Dkin} which
contains rational functions and HPLs up to weight 4,
and perform a series expansion in $T$ to a finite power $n$.
Then we expand Eqs.~\eqref{eq:resType1} and \eqref{eq:resType2} up to $\ord(T^n)$
by taking residues, and compute the remaining one-dimensional MB integrals.
The result can then be fitted to the series expansion of the ansatz;
the fitting procedure consists
of solving a system of linear equations to determine the unknown coefficients
of the ansatz.

An ansatz which includes weight 4 functions is rather large, requiring a series
expansion to a high power $n$ to completely fix its coefficients. In practice,
our experience shows that the arc does not contribute to the
higher-transcendental-weight contributions, allowing us to limit the size of
the ansatz and thus the required depth of the series expansions.

In the following, we demonstrate this idea with an explicit example that
is present in our calculation. We have a two-dimensional MB expression $I$
and perform the residue summation as described above. This leads to
\begin{eqnarray}
\label{eq:sum2fit}
I_{\mathrm{sum}} &=&
 \bigg( 
-\frac{4}{(x+1)^2}-\frac{12 x}{(x+1)^2}-\frac{12 x^2}{(x+1)^2}-\frac{4 x^3}{(x+1)^2}-\frac{6 H(-1,x)}{(x+1)^2}-\frac{4 H(-1,x)}{x (x+1)^2} \nonumber \\[1mm] 
&&-\, \frac{2 x H(-1,x)}{(x+1)^2}
 \bigg) \, \log(x) + I_{\mathrm{sum}}^{(\mathrm{high})} \,, 
\end{eqnarray}
where
\begin{eqnarray}
I_{\mathrm{sum}}^{(\mathrm{high})} &=& \bigg( - \frac{20 x^3 H(0,-1,x)}{(x+1)^2}-\frac{56 x^2 H(0,-1,x)}{(x+1)^2}-\frac{38 x H(0,-1,x)}{(x+1)^2} +\frac{4 H(0,-1,x)}{(x+1)^2} \nonumber \\[1mm] 
&&+\, \frac{8 H(0,-1,x)}{x (x+1)^2}  + \; \text{weight 3 transcendental functions}
 \bigg) \, \log(x) \,,
\end{eqnarray}
and $x=T/S$. $I_{\mathrm{sum}}^{(\mathrm{high})}$ contains functions of
transcendental weight 3 and 4 which, in our calculation, are
correctly computed by the residue sums. This
suggests an ansatz which contains undetermined coefficients in front of
functions only up to transcendental weight 2. Here we choose 
\begin{eqnarray}
\label{eq:ansatz2fit}
I_{\mathrm{ansatz}} &=&  \bigg( \frac{c_1}{(x+1)^2} +\frac{c_2 x}{(x+1)^2} +\frac{c_3 x^2}{(x+1)^2} + \frac{c_4 x^3}{(x+1)^2} +\frac{c_5 H(-1,x)}{x (x+1)^2} +\frac{c_6 H(-1,x)}{(x+1)^2} \nonumber \\[1mm]
&& +\, \frac{c_7 x H(-1,x)}{(x+1)^2}+\frac{c_8 x^2 H(-1,x)}{(x+1)^2} + \frac{c_9 x^3 H(-1,x)}{(x+1)^2} \bigg)\,\log(x) + I_{\mathrm{sum}}^{(\mathrm{high})} \,,
\end{eqnarray}
with the nine free parameters $c_1,\ldots, c_9$.

Using numerical integration and the \texttt{PSLQ} algorithm we can
construct a series expansion of $I$ which is given by
\begin{eqnarray}
\label{eq:exp2fit}
I_{\mathrm{exp}}&=&
\bigg( 4 -40 x -\frac{112 x^2}{9} -\frac{1123 x^3}{108} +\frac{148453 x^4}{5400} -\frac{2409487 x^5}{54000} +\frac{82787909 x^6}{1323000}  \nonumber \\[1mm]
&& - \, \frac{3017222321 x^7}{37044000} +  \frac{22492195259 x^8}{222264000} -\frac{487063561297 x^9}{4000752000} + \frac{1730875605497 x^{10}}{12102274800} \nonumber \\[1mm]
&& + \, \ord(x^{11}) \bigg) \, \log(x) \,.
\end{eqnarray}
Note that here the arc contributions are included correctly.
By performing a series expansion of Eq.~\eqref{eq:ansatz2fit} and comparing to Eq.~\eqref{eq:exp2fit}
we obtain an over-determined system of linear equations with the solution
\begin{eqnarray}
\{c_1 = -4, \, c_2 = -12, \, c_3 = -12,\, c_4 = -4, \, c_5 = 0,\, c_6 = 0, \, c_7 = 0, \, c_8 = 0, \, c_9 = 0\}\,.
\end{eqnarray}
After inserting the coefficients into Eq.~\eqref{eq:ansatz2fit} we finally obtain the
true result for $I$ which replaces Eq.~(\ref{eq:sum2fit}).

\paragraph{Results:} After solving all MB integrals and adding the result from
the hard region, we derive the final solution of this ``7+2''-line master
integral 
\begin{eqnarray}
\mathcal{I}_{7,2} \! &=\!&
\frac{T}{4 S^2}H\!\!\left(0,\tfrac{m_t^2}{S}\right)^4  +\frac{1}{T} \, \bigg\{ H\!\!\left(0,\tfrac{T}{S}\right) H\!\!\left(0,\tfrac{m_t^2}{S}\right)^3+\left(-\frac{3}{2} H\!\!\left(0,\tfrac{T}{S}\right)^2-\frac{\pi ^2}{2}\right) H\!\!\left(0,\tfrac{m_t^2}{S}\right)^2 \vphantom{\bigg\{}\nonumber\\[-1mm]
&&{}\hspace{-5mm} +\,\left[H\!\!\left(0,\tfrac{T}{S}\right)^3+\left(6 H\!\!\left(0,-1,\tfrac{T}{S}\right)+\pi ^2\right) H\!\!\left(0,\tfrac{T}{S}\right)-6 H\!\!\left(0,0,-1,\tfrac{T}{S}\right)\right] H\!\!\left(0,\tfrac{m_t^2}{S}\right) \vphantom{\bigg\{}\nonumber\\[-1mm]
&&{}\hspace{-5mm} +\, H\!\!\left(-1,\tfrac{T}{S}\right) \bigg[\left(-3 H\!\!\left(0,\tfrac{T}{S}\right)^2-3 \pi ^2\right) H\!\!\left(0,\tfrac{m_t^2}{S}\right)+2 H\!\!\left(0,\tfrac{T}{S}\right)^3 \vphantom{\bigg\{}\nonumber\\[-1mm]
&&{}\hspace{-5mm} +\, \left(8 H\!\!\left(0,-1,\tfrac{T}{S}\right)+\frac{10 \pi ^2}{3}\right) H\!\!\left(0,\tfrac{T}{S}\right)-8 H\!\!\left(0,0,-1,\tfrac{T}{S}\right)\bigg]
-\frac{1}{4} H\!\!\left(0,\tfrac{m_t^2}{S}\right){}^4 \vphantom{\bigg\{}\nonumber\\[-1mm]
&&{}\hspace{-5mm} +\,\zeta (3) \left(8-8 H\!\!\left(-1,\tfrac{T}{S}\right)\right)-\frac{1}{4} H\!\!\left(0,\tfrac{T}{S}\right)^4+\left(-3 H\!\!\left(0,-1,\tfrac{T}{S}\right)-\frac{\pi ^2}{2}\right) H\!\!\left(0,\tfrac{T}{S}\right)^2 \vphantom{\bigg\{}\nonumber\\[-1mm]
&&{}\hspace{-5mm} -\,8 H\!\!\left(0,-1,-1,\tfrac{T}{S}\right) H\!\!\left(0,\tfrac{T}{S}\right)+H\!\!\left(-1,\tfrac{T}{S}\right)^2 \left(-2 H\!\!\left(0,\tfrac{T}{S}\right)^2-2 \pi ^2\right)-\frac{1}{3} \pi ^2 H\!\!\left(0,-1,\tfrac{T}{S}\right) \vphantom{\bigg\{}\nonumber\\[-1mm]
&&{}\hspace{-5mm} +\,8 H\!\!\left(0,0,-1,-1,\tfrac{T}{S}\right)+6 H\!\!\left(0,0,0,-1,\tfrac{T}{S}\right)-\frac{5}{3}-\frac{\pi ^2}{18}-\frac{19 \pi ^4}{180} + C_{T} \bigg\} \vphantom{\bigg\{}\nonumber\\[-1mm]
&&{}\hspace{-5mm}  +\, \frac{1}{S} \, \bigg\{ \zeta (3) \left(10 H\!\!\left(0,\tfrac{m_t^2}{S}\right)-8 H\!\!\left(-1,\tfrac{T}{S}\right)-4 H\!\!\left(0,\tfrac{T}{S}\right)+18\right)+H\!\!\left(0,\tfrac{T}{S}\right) H\!\!\left(0,\tfrac{m_t^2}{S}\right)^3 \vphantom{\bigg\{}\nonumber\\[-1mm]
&&{}\hspace{-5mm} -\, \left(2 H\!\!\left(0,\tfrac{T}{S}\right)^2 + \frac{2 \pi ^2}{3}\right) H\!\!\left(0,\tfrac{m_t^2}{S}\right)^2+\bigg[\frac{5}{3} H\!\!\left(0,\tfrac{T}{S}\right)^3+\left(6 H\!\!\left(0,-1,\tfrac{T}{S}\right)+\frac{5 \pi ^2}{3}\right) H\!\!\left(0,\tfrac{T}{S}\right) \vphantom{\bigg\{}\nonumber\\[-1mm]
&&{}\hspace{-5mm} -\,6 H\!\!\left(0,0,-1,\tfrac{T}{S}\right)\bigg] H\!\!\left(0,\tfrac{m_t^2}{S}\right)+H\!\!\left(-1,\tfrac{T}{S}\right) \bigg[\left(-3 H\!\!\left(0,\tfrac{T}{S}\right)^2-3 \pi ^2\right) H\!\!\left(0,\tfrac{m_t^2}{S}\right) \vphantom{\bigg\{}\nonumber\\[-1mm]
&&{}\hspace{-5mm} +\,2 H\!\!\left(0,\tfrac{T}{S}\right)^3+\left(8 H\!\!\left(0,-1,\tfrac{T}{S}\right)+\frac{10 \pi ^2}{3}\right) H\!\!\left(0,\tfrac{T}{S}\right)-8 H\!\!\left(0,0,-1,\tfrac{T}{S}\right)\bigg] \vphantom{\bigg\{}\nonumber\\[-1mm]
&&{}\hspace{-5mm}  +\, \frac{1}{24} H\!\!\left(0,\tfrac{m_t^2}{S}\right)^4 -  \frac{1}{2} H\!\!\left(0,\tfrac{T}{S}\right)^4 - \left(2 H\!\!\left(0,-1,\tfrac{T}{S}\right) + \pi ^2\right) H\!\!\left(0,\tfrac{T}{S}\right)^2  + \frac{2}{3} \pi ^2 H\!\!\left(0,-1,\tfrac{T}{S}\right) \vphantom{\bigg\{}\nonumber\\[-1mm]
&&{}\hspace{-5mm}  -\, \left(8 H\!\!\left(0,-1,-1,\tfrac{T}{S}\right) + 4 H\!\!\left(0,0,-1,\tfrac{T}{S}\right)\right) H\!\!\left(0,\tfrac{T}{S}\right)  - H\!\!\left(-1,\tfrac{T}{S}\right)^2 \left(2 H\!\!\left(0,\tfrac{T}{S}\right)^2 + 2 \pi ^2\right) \vphantom{\bigg\{}\nonumber\\[-1mm]
&&{}\hspace{-5mm} +8 H\!\!\left(0,0,-1,-1,\tfrac{T}{S}\right)+12
   H\!\!\left(0,0,0,-1,\tfrac{T}{S}\right) +\frac{5 \pi
   ^2}{18}-\frac{2}{3}-\frac{2 \pi ^4}{9} + C_{S} \bigg\} + \ord(\epsilon
   ,m_t^2) \,, \vphantom{\bigg\{}\nonumber\\[-1mm]
\end{eqnarray}
where the constants $C_{T}$ and $C_{S}$ originate from three-dimensional
MB integrals  which are discussed in Appendix~\ref{app:MBconst}.


\subsubsection{Crossing and analytic continuation}

As stated above, we only calculate the boundary conditions for the subset of master integrals
for which the Euclidean region is defined for $S,T>0$ and $U<0$.
The boundary conditions for all other master integrals can be obtained by applying one
of the five crossing relations:
\begin{eqnarray}
    T \rightarrow U \ ;\quad \ S \rightarrow U \ ;\quad \ S \leftrightarrow T \ ;\quad \ 
    T \rightarrow U \, , \, S \rightarrow T \ ;\quad \
    T \rightarrow S \, , \, S \rightarrow U \ .
\end{eqnarray}
While the rational dependence can be easily obtained via these replacements,
the HPLs need analytic continuation.

Due to our choice of the Euclidean region we start with HPLs of the argument
$x=T/S$, which are real in this region.  To analytically continue to the
physical region, we have to arrive at the argument
$x^\prime = - T/S = T/s = -x$.  The transformation of HPLs to the negative
argument is implemented in \texttt{HarmonicSums} and \texttt{HPL}.  However, we
have to take care to use the correct sign for the analytic continuation.  We
have $s = s + {\rm i} \, \varepsilon$, so $x = x + {\rm i} \, \varepsilon$ and
therefore have to use the `$+$' sign for the analytic continuation which leads to
\begin{eqnarray}
    H(0,x) &=& H(0,x^\prime) + {\rm i} \, \pi \,.
\end{eqnarray}
Using \texttt{HarmonicSums} or \texttt{HPL} we can transform the argument
of all occurring HPLs to the physical region. For example, we have
\begin{eqnarray}
    H(0,-1,x) &=& -H(0,1,x^\prime) , \nonumber\\ 
    H(0,-1,-1,x) &=& H(0,1,1,x^\prime) 
    ~.
\end{eqnarray}

The analytic continuation of the HPLs after the application of the different
crossings can be obtained in a similar manner, but require more involved
transformations.  For example, after the crossing $T \rightarrow U$ we end up
with HPLs of the argument $y=-( 1 + T/S + {\rm i} \, \varepsilon )$.  We can
map these HPLs back to argument $x^\prime$ by first applying the
transformation $y \to -y = y^\prime$ and afterwards
$y^\prime \to 1 - y^\prime = x^\prime$.  The sign for the analytic
continuation has to be chosen as `$-$' for the first and `$+$' for the second
transformation.  This results, for example, in 
\begin{eqnarray}
    H(0,-1,y) &=& H(0,1,x^\prime)-H(0,x^\prime)H(1,x^\prime)-\zeta(2) , \nonumber\\ 
    H(0,-1,-1,y) &=& -H(0,0,1,x^\prime)+ H(0,x^\prime)H(0,1,x^\prime)
    \\ \nonumber &&
        -\frac{1}{2}H(0,x^\prime)^2 H(1,x^\prime) + \zeta(3)
    ~.
\end{eqnarray}

As a final example, let us look at the crossing $S \leftrightarrow T$.  Here, we
find HPLs of argument $w = S/T - {\rm i} \, \varepsilon$.  We can map these
HPLs back to argument $x^\prime$ by first applying the transformation
$w \to 1/w = x$ and afterwards continue as for the first example.  We find
\begin{eqnarray}
    H(0,-1,w) &=& H(0,1,x^\prime) + \frac{1}{2} H(0,x^\prime)^2 - \frac{\pi^2}{3} + {\rm i} \, \pi \, H(0,x^\prime)  , \nonumber\\ 
    H(0,-1,-1,w) &=& - H(0,1,1,x^\prime) + H(0,0,1,x^\prime)
    - H(0,x^\prime) H(0,1,x^\prime) - \frac{1}{6} H(0,x^\prime)^3
    \nonumber \\ &&
    + \frac{\pi^2}{2} H(0,x^\prime) + \zeta(3)
    - {\rm i} \, \pi \biggl(
    H(0,1,x^\prime) + \frac{1}{2} H(0,x^\prime)^2 - \frac{\pi^2}{6}
    \biggr)
    ~.
\end{eqnarray}

The analytic continuation for the other crossings can be derived analogously.
In total we can express all 140 master integrals through the following set of 
HPLs:
\begin{eqnarray}
    H\!\!\left(0, \tfrac{T}{s} \right) &=& \log\left( \tfrac{T}{s} \right) , \nonumber\\
    H\!\!\left(1, \tfrac{T}{s} \right) &=& - \log\left( 1 - \tfrac{T}{s} \right) , \nonumber\\
    H\!\!\left(0, 1, \tfrac{T}{s} \right) &=& \text{Li}_{2}\left( \tfrac{T}{s} \right), \nonumber\\ 
    H\!\!\left(0, 0, 1, \tfrac{T}{s} \right) &=& \text{Li}_{3}\left( \tfrac{T}{s} \right) , \nonumber\\
    H\!\!\left(0, 1, 1, \tfrac{T}{s} \right) &=& -\text{Li}_3\left(1-\tfrac{T}{s}\right)+\text{Li}_2\left(1-\tfrac{T}{s}\right) \log\left(1-\tfrac{T}{s}\right)
    \nonumber \\ &&
    +\tfrac{1}{2} \log \left(\tfrac{T}{s}\right) \log^2\left(1-\tfrac{T}{s}\right)+\zeta(3), \nonumber\\
    H\!\!\left(0, 0, 0, 1, \tfrac{T}{s} \right) &=& \text{Li}_{4}\left( \tfrac{T}{s} \right), \nonumber\\ 
    H\!\!\left(0, 0, 1, 1, \tfrac{T}{s} \right) &=& \text{S}_{2,2}\left( \tfrac{T}{s} \right) , \nonumber\\
    H\!\!\left(0, 1, 1, 1, \tfrac{T}{s} \right) &=& -\text{Li}_4\left(1-\tfrac{T}{s}\right)-\tfrac{1}{2} \text{Li}_2\left(1-\tfrac{T}{s}\right) \log ^2\left(1-\tfrac{T}{s}\right) +\tfrac{\pi ^4}{90}
    \nonumber \\ &&
    +\text{Li}_3\left(1-\tfrac{T}{s}\right) \log \left(1-\tfrac{T}{s}\right)-\tfrac{1}{6} \log \left(\tfrac{T}{s}\right) \log ^3\left(1-\tfrac{T}{s}\right) .
\end{eqnarray} 
While the expression in terms of HPLs is more convenient for analytic
manipulations, the expressions in terms of polylogarithms ($\text{Li}_n(x)$)
and Nielsen polylogarithms ($S_{n,m}(x)$) might be more convenient for
numerical evaluations, since many standard math libraries already contain
implementations.


In the supplementary material to this paper~\cite{progdata} we provide
the analytic results for all 140 master integrals.



\section{\label{sec::FF}Form factors for $gg\to HH$}

The contribution to the form factors of $gg\to HH$ from diagrams
of Fig.~\ref{fig::diags1} is infrared finite and has only ultraviolet
divergences. They are removed by renormalizing the top quark mass
and Yukawa coupling in the leading order contributions.
The counterterms are well known in the on-shell scheme, see, e.g.,
Ref.~\cite{Denner:1991kt}. In this work it is sufficient to perform the
renormalization in the $\overline{\rm MS}$ scheme. The corresponding
mass counterterm is given by (see, e.g., Eq.~(31) of Ref.~\cite{Kniehl:2004hfa})
\begin{eqnarray}
  m_t^0 = \overline{m}_t\left[
  1 + \frac{\alpha}{\pi s_W^2 \epsilon}\left(
  \frac{3}{32} \frac{\overline{m}_t^2}{m_W^2}
  + \frac{N_C}{4} \frac{\overline{m}_t^4}{m_W^2 m_H^2}
  \right)
  \right]
  = \overline{m}_t\left[
  1 + \frac{\alpha_t}{\pi\epsilon}
  \left(\frac{3}{16} + \frac{N_C}{2}\frac{\overline{m}_t^2}{m_H^2}\right)
  \right]
  \,,
\end{eqnarray}
where $\alpha$ is the fine structure constant, $s_W\equiv \sin\theta_W$ is the
sine of the weak mixing angle and $N_C=3$. The second term inside the round
brackets originates from the tadpole contribution\footnote{For a recent
  improved prescription for the renormalization of tadpole contributions we
  refer to~\cite{Dittmaier:2022maf}.} and is only provided for completeness;
it is not used in this paper.

The finite form factors $F_{\rm box1}$ and $F_{\rm box2}$ are expanded up to
$(m_t^2)^{57}$ and $m_H^4$ in approach~(A) and up to $(m_t^2)^{58}$,
$(m_H^{ext})^4$ and $\delta^3$ in approach~(B).
Note that one factor $m_t^2$ is collected in $\alpha_t$ (see
Eq.~(\ref{eq::alphat})) such that the expansion up to $(m_t^2)^{56}$
and $(m_t^2)^{57}$ are available for the Pad\'e method.
We follow Ref.~\cite{Davies:2020lpf} and construct the so-called
``pole distance re-weighted'' Pad\'e approximants and the corresponding
uncertainties (see Section~4 of~\cite{Davies:2020lpf} for a detailed
discussion), in which Pad\'e approximants $[n/m]$ are included which satisfy
\begin{eqnarray}
  	N_{\rm low}\le n+m \le N_{\rm high}\quad
	\textnormal{and}\quad N_{\rm low} \le n + m - | n - m | \,.
	\label{eq::N_low_high}
\end{eqnarray}
For approach~(A) we choose $\{N_{\rm low},N_{\rm high}\}=\{50,56\}$ and 
for approach~(B) $\{N_{\rm low},N_{\rm high}\}=\{51,57\}$.
Note that in~\cite{Davies:2019dfy} only terms up to $(m_t^2)^{16}$ are available. We observe
that including more $m_t$ expansion terms in the construction of the
Pad\'e approximations leads to a significant stabilization of the results, in
particular for lower values of $p_T$.
For the numerical analyses we choose $m_t=173$~GeV,
$m_H=125$~GeV and set $\mu^2=s$.

\begin{figure}[t]
  \centering
  \includegraphics[width=0.4\textwidth]{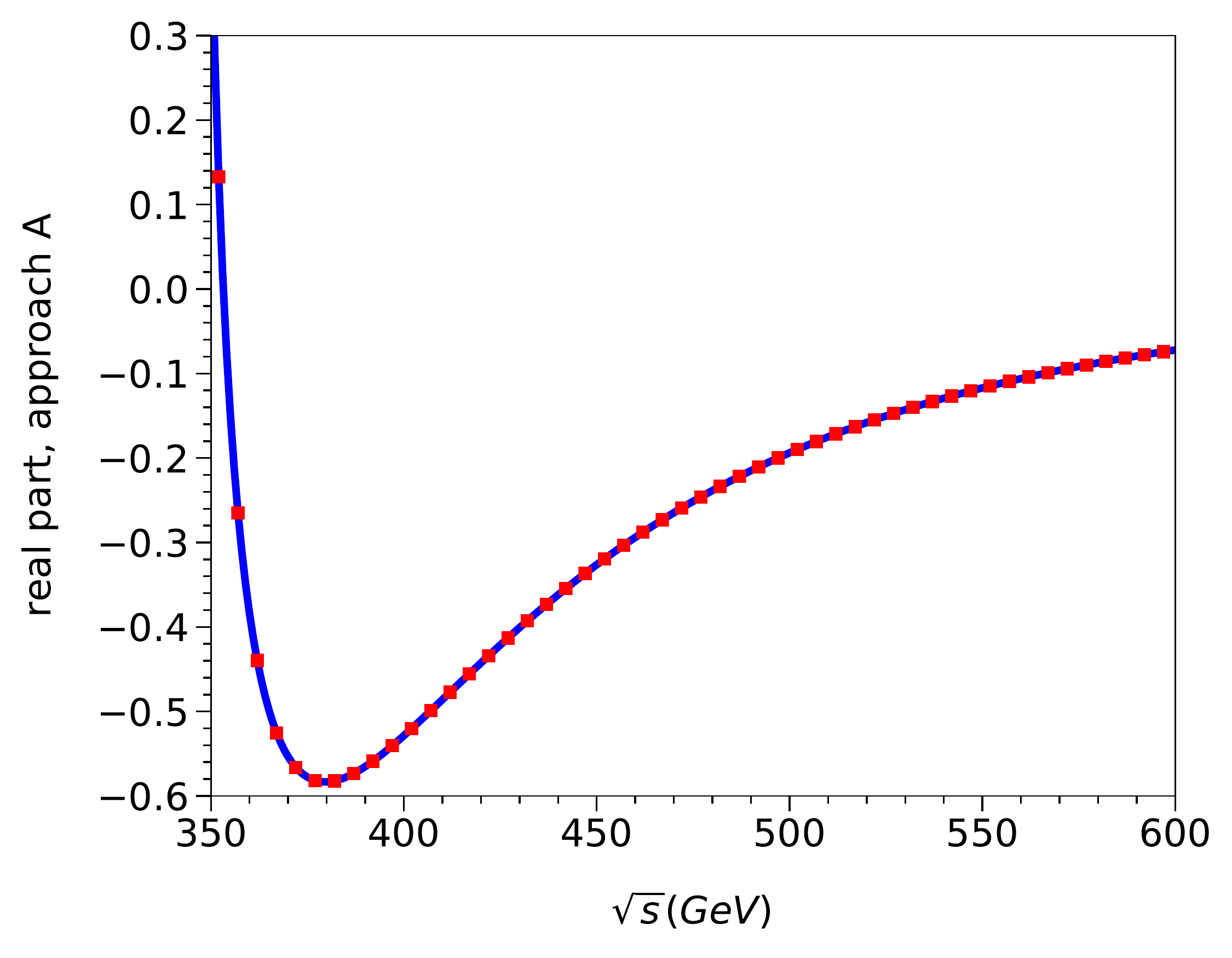}
  \includegraphics[width=0.4\textwidth]{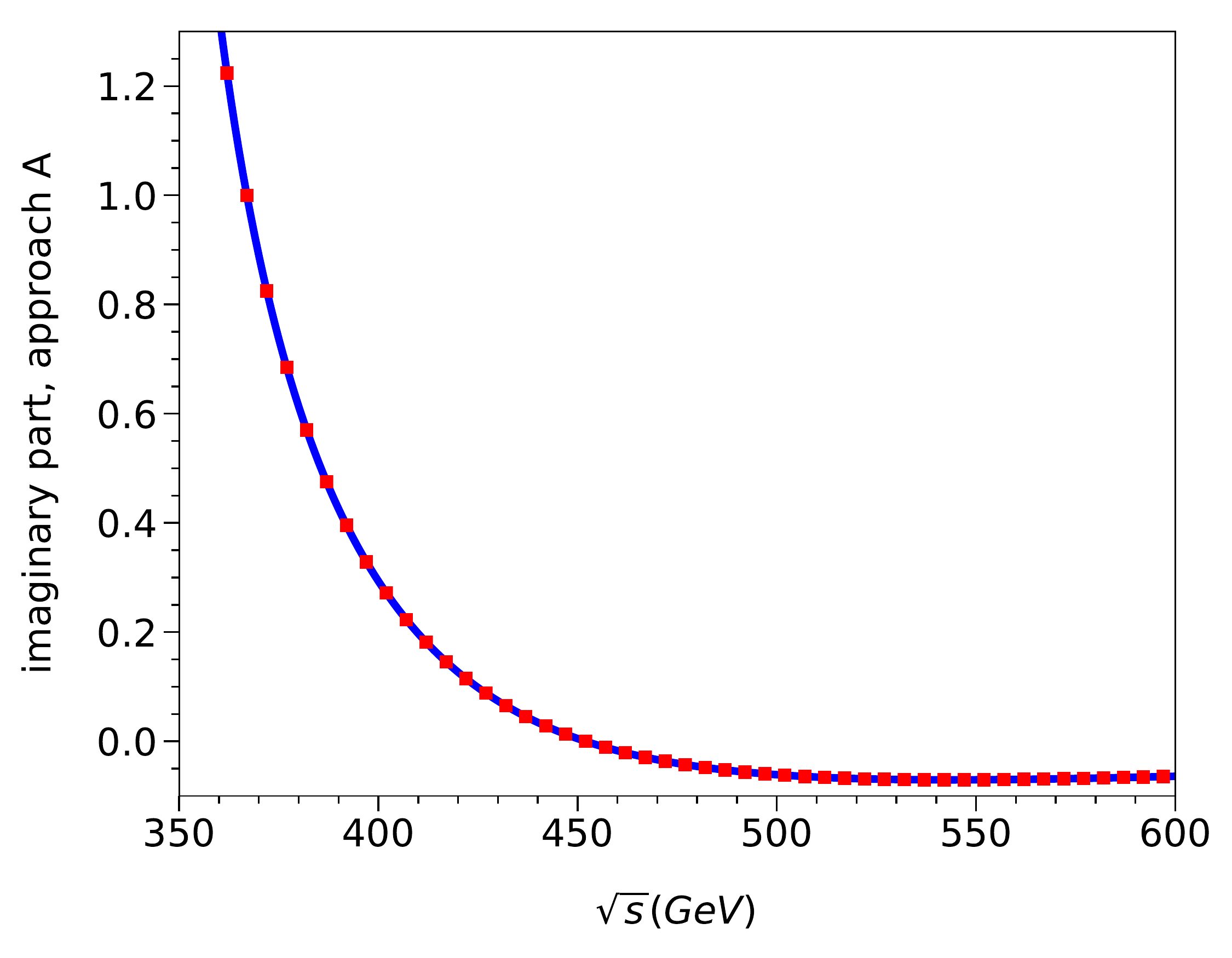}
  \\ (a) \\
  \includegraphics[width=0.4\textwidth]{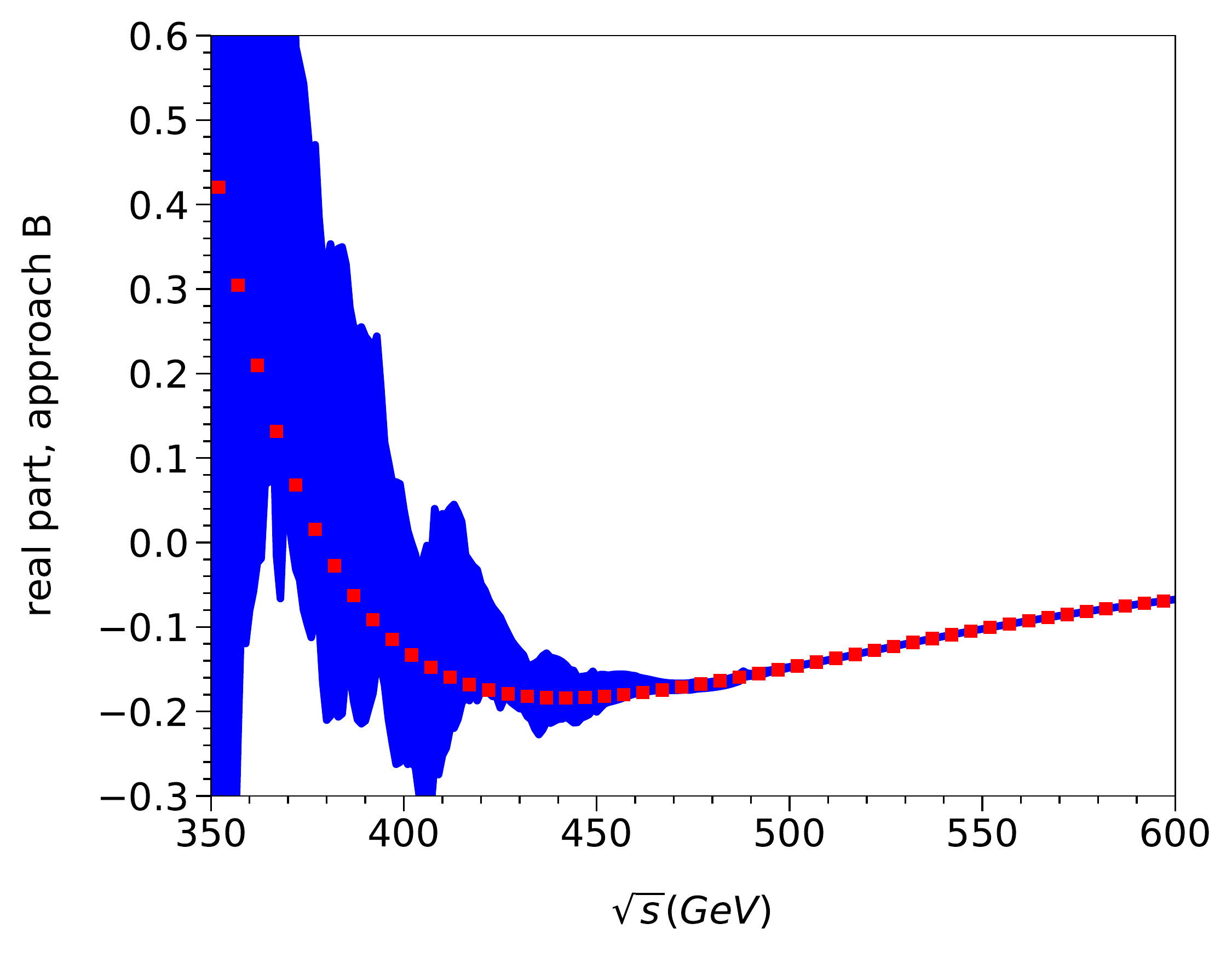}
  \includegraphics[width=0.4\textwidth]{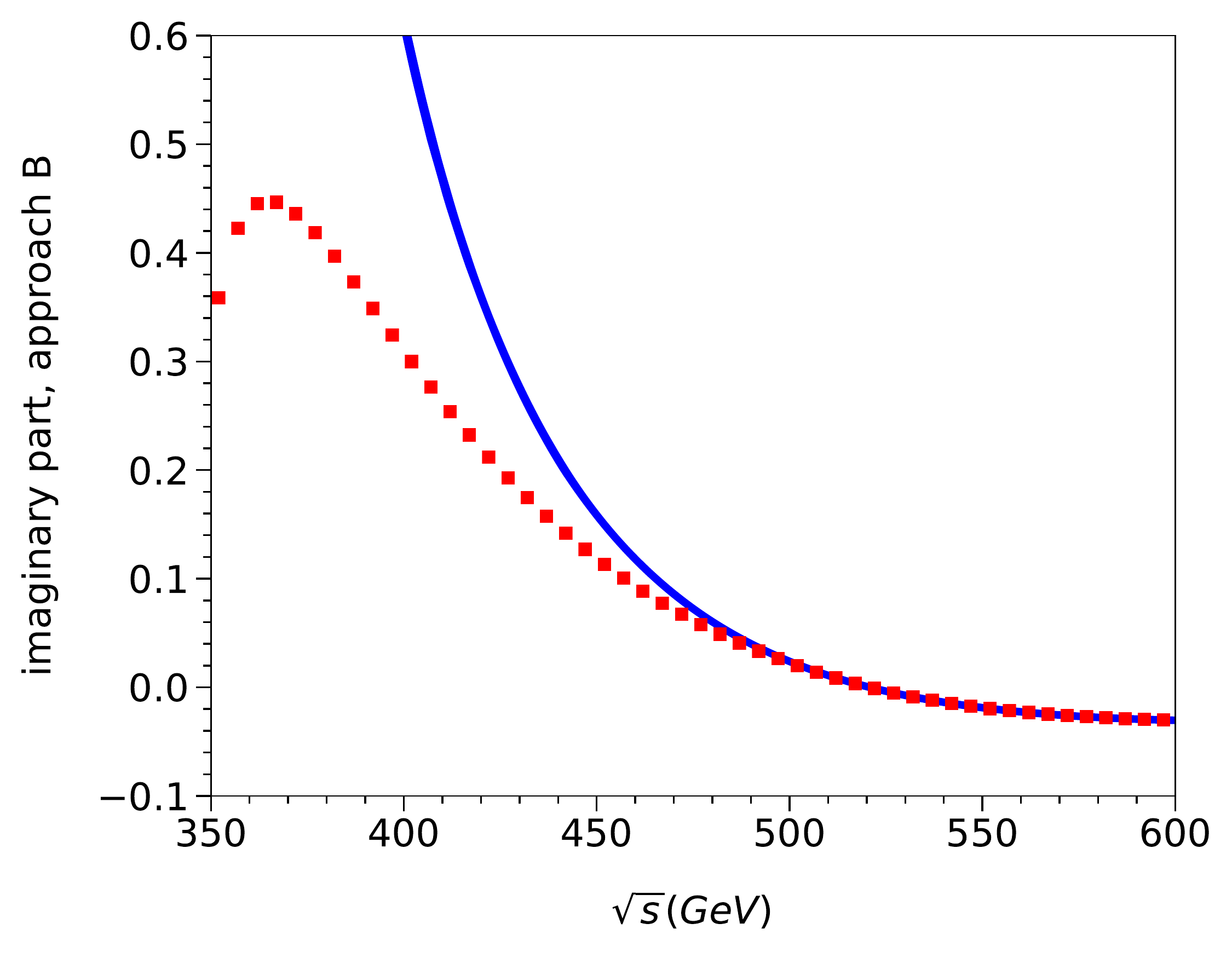}
  \\ (b) \\
  \caption{\label{fig::G4}Real and imaginary parts of the seven-line master
    integral (see Fig.~\ref{fig:DBox}) for $p_T=120$~GeV. 
    Top: the line in the middle is massless;
    bottom: all lines are massive.
    Numerical results from {\tt FIESTA}
    are shown in red; Pad\'e results together with the corresponding
    uncertainty band are shown in blue. }
\end{figure}

Before discussing the results for the physical form factors we apply our
approximation method to the seven-line double box integral (see Fig.~\ref{fig:DBox})
  where all internal lines are
  massive.  This is one of our master integrals, which we have expanded up to
  $(m_t^2)^{60}$.  For this integral it is possible to obtain precise numerical
  results using {\tt FIESTA}.  In Fig.~\ref{fig::G4}(b) we compare, for
  $p_T=120$~GeV, the real and imaginary parts of the Pad\'e method to the
  numerical results. For the Pad\'e method we use
  $\{N_{\rm low},N_{\rm high}\}=\{50,60\}$, the same choice as we make for the form factors.
  For values of $\sqrt{s}\approx 500$~GeV and higher the Pad\'e uncertainties are
  very small and we find perfect agreement between the Pad\'e and {\tt FIESTA}
  results.  For lower $\sqrt{s}$ the Pad\'e uncertainties in the real part
  grow. It is nevertheless interesting to see that the central values are
  close to the numerical results.  On the contrary, for the imaginary part the
  Pad\'e uncertainties remain small but there is a clear deviation from the
  exact result.  This can be explained as follows: The integral we consider admits
  two- and three-particle cuts. For the latter
  $\sqrt{s} = 3 m_t = 519$~GeV which is about the starting point for the
  deviations; the Pad\'e method is not expected to be able to
  approximate the exact function below the cut, which we clearly see in the
  imaginary part in Fig.~\ref{fig::G4}(b).

  In Fig.~\ref{fig::G4}(a) we show the analogous result for the seven-line
  master integral of approach~(A) where middle line is massless. This
  integral only has cuts through two massive lines (and possibly also
  a massless line) and indeed, we observe good
  agreement of the Pad\'e and {\tt FIESTA} results, even close to the
  top quark pair threshold at $2 m_t = 346$~GeV.

\begin{figure}[t]
  \centering
  \begin{tabular}{cc}
    \includegraphics[width=0.45\textwidth]{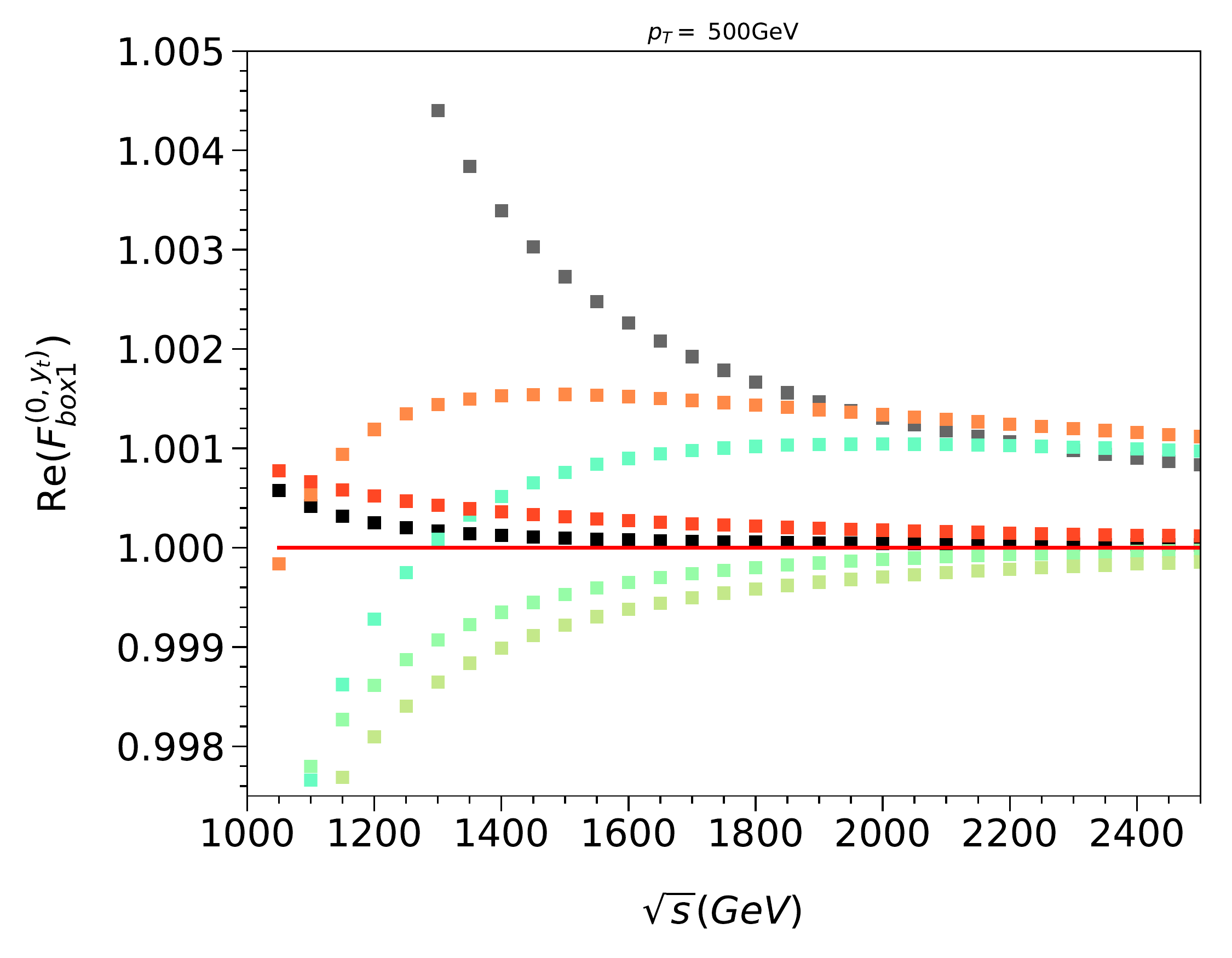}
    &
    \includegraphics[width=0.45\textwidth]{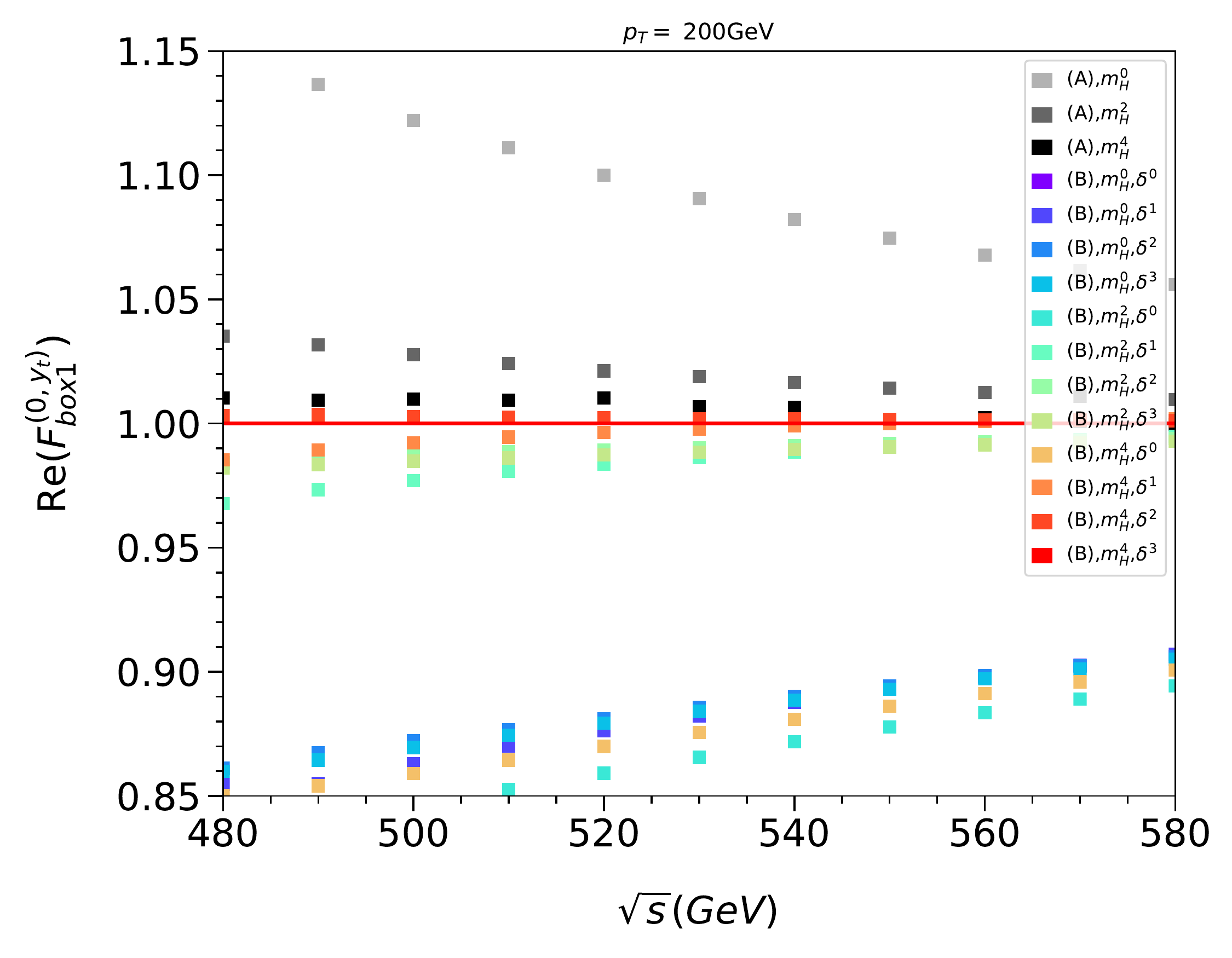}
      \\
    (a) & (b)
  \end{tabular}
  \caption{\label{fig::F12_ratio}Real part of $F_{\rm box1}$ for different
    values of $p_T$ and various expansion terms in
    $m_H$ and $\delta$.}
\end{figure}

Let us now move to the form factors $F_{\rm box1}$ and $F_{\rm box2}$
and  discuss the quality of the expansions in $m_H$ and $\delta$.
For this purpose we fix $p_T$ and plot various different depths.
We normalize all curves to the highest-available depth of approach (B),
which includes $m_H^4$ and $\delta^3$.

In Fig.~\ref{fig::F12_ratio}(a) the result is shown for the real part of
$F_{\rm box1}$ for $p_T=500$~GeV. The colours correspond to
approach~(B) and the results from approach~(A) are shown in gray and black.
The $y$ axis spans a range below 1\% and all approximations
which include at least $m_H^2$ terms in approach~(A) and
$m_H^2$ and $\delta^1$ terms in approach~(B) are visible in the plot
and thus show a deviation well below the percent level.

In Fig.~\ref{fig::F12_ratio}(b) we show results for 
$p_T=200$~GeV and $\sqrt{s}$ values between 480~GeV and 
580~GeV.
For larger values of $\sqrt{s}$ the form factor crosses zero and 
the ratios inflate. Beyond the zero crossing the ratios are a
similar size to those of Fig.~\ref{fig::F12_ratio}(a).
The result from approach~(A) show a deviation of about 10\% in case the Higgs
mass is neglected. It reduced to below 5\% after including the $m_H^2$ terms
and is of order 1\% after including also the quartic terms.
The situation is similar for approach~(B): Once quadratic terms in $m_H$ and
$\delta$ are include the deviation from 1 is below 5\%. Including more
expansion terms in $m_H$ and $\delta$ further stabilizes the approximations.

We conclude that the inclusion of the quartic terms in $m_H$ and
cubic terms in $\delta$ provides an approximation to the (unknown)
exact result below the percent level (see also Fig.~2 or Ref.~\cite{Chen:2022rua}
which shows a comparison for $gg\to ZH$).

\begin{figure}[t]
  \centering
  \begin{tabular}{cc}
    \includegraphics[width=0.45\textwidth]{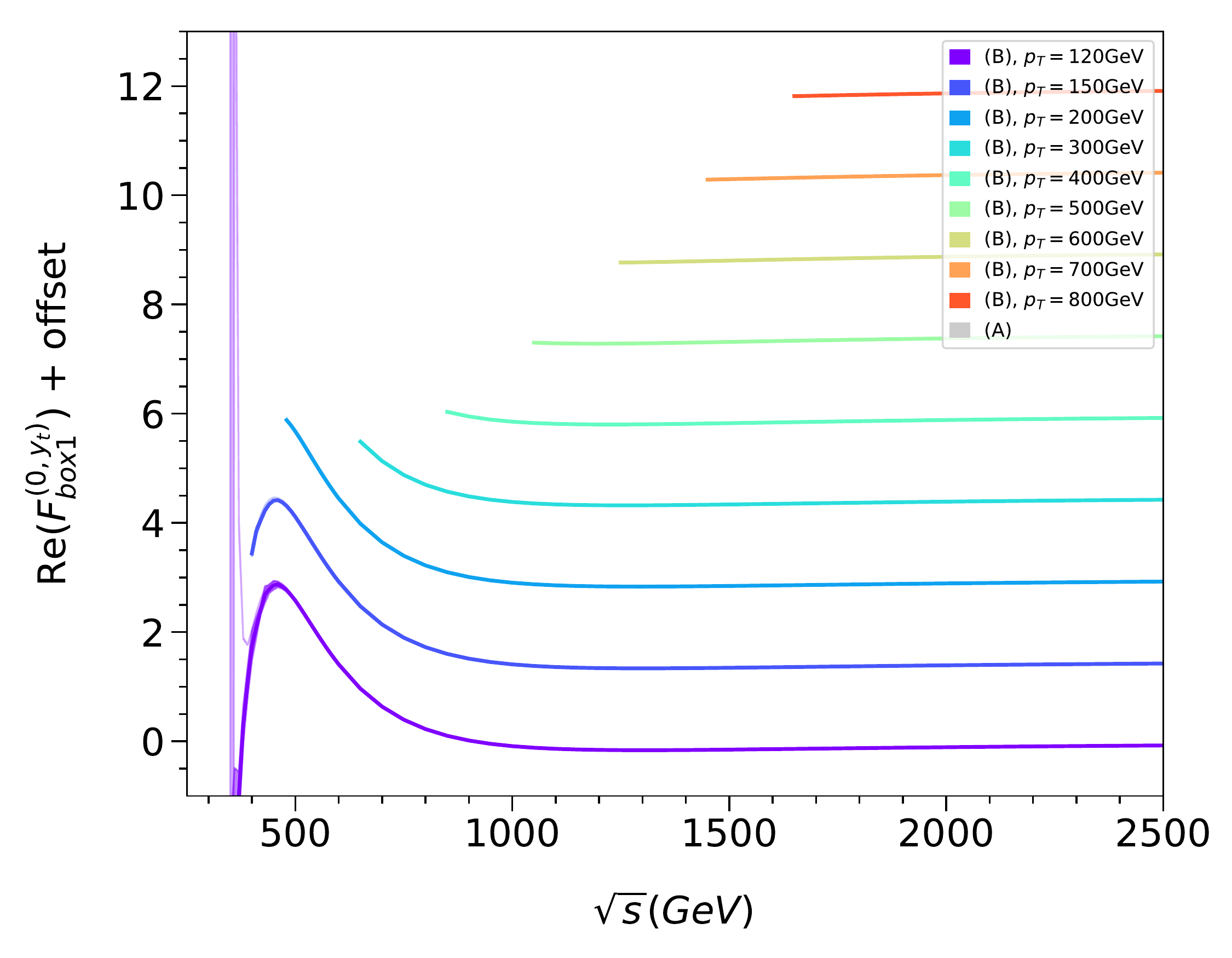}
    &
    \includegraphics[width=0.45\textwidth]{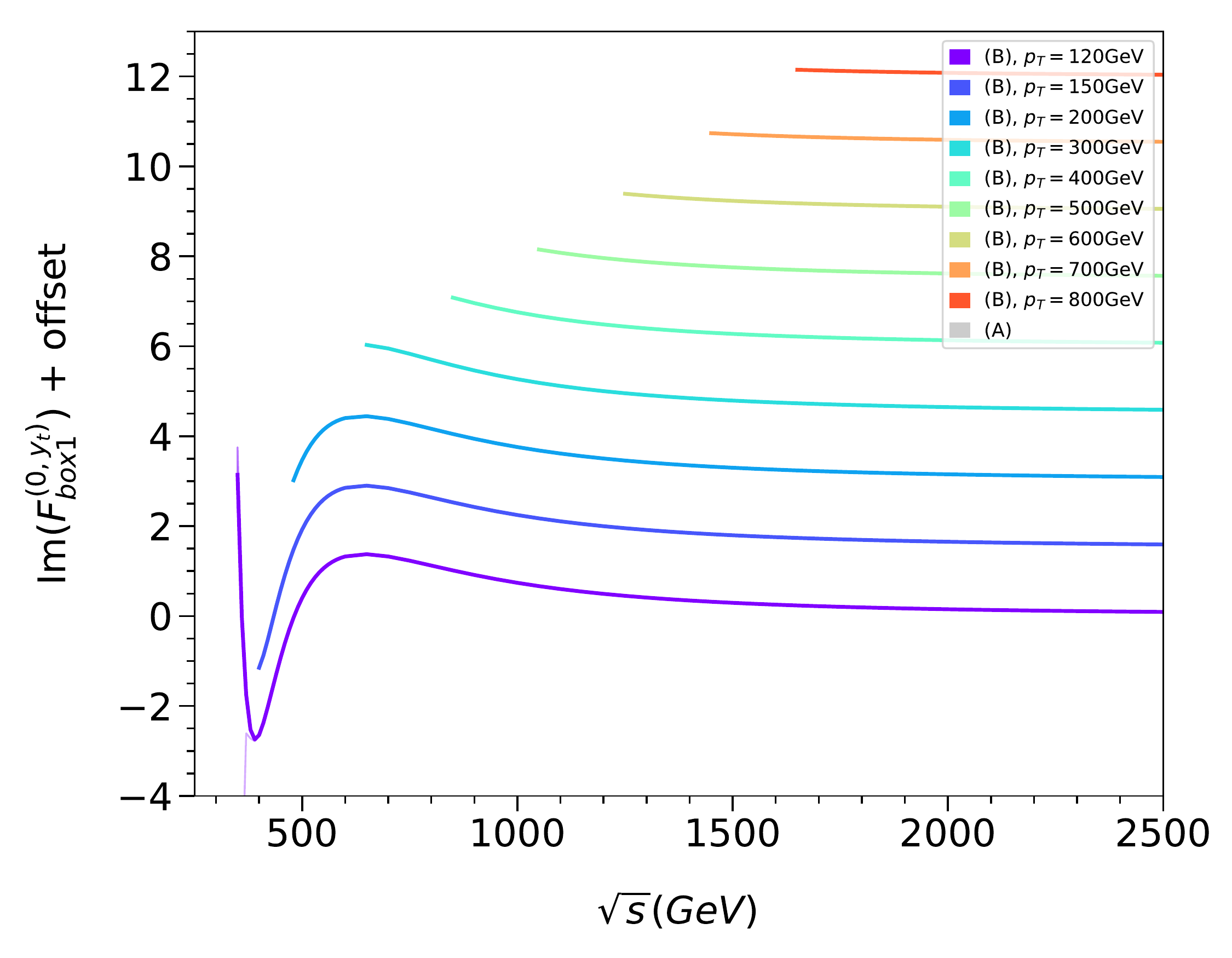}
  \\
    \includegraphics[width=0.45\textwidth]{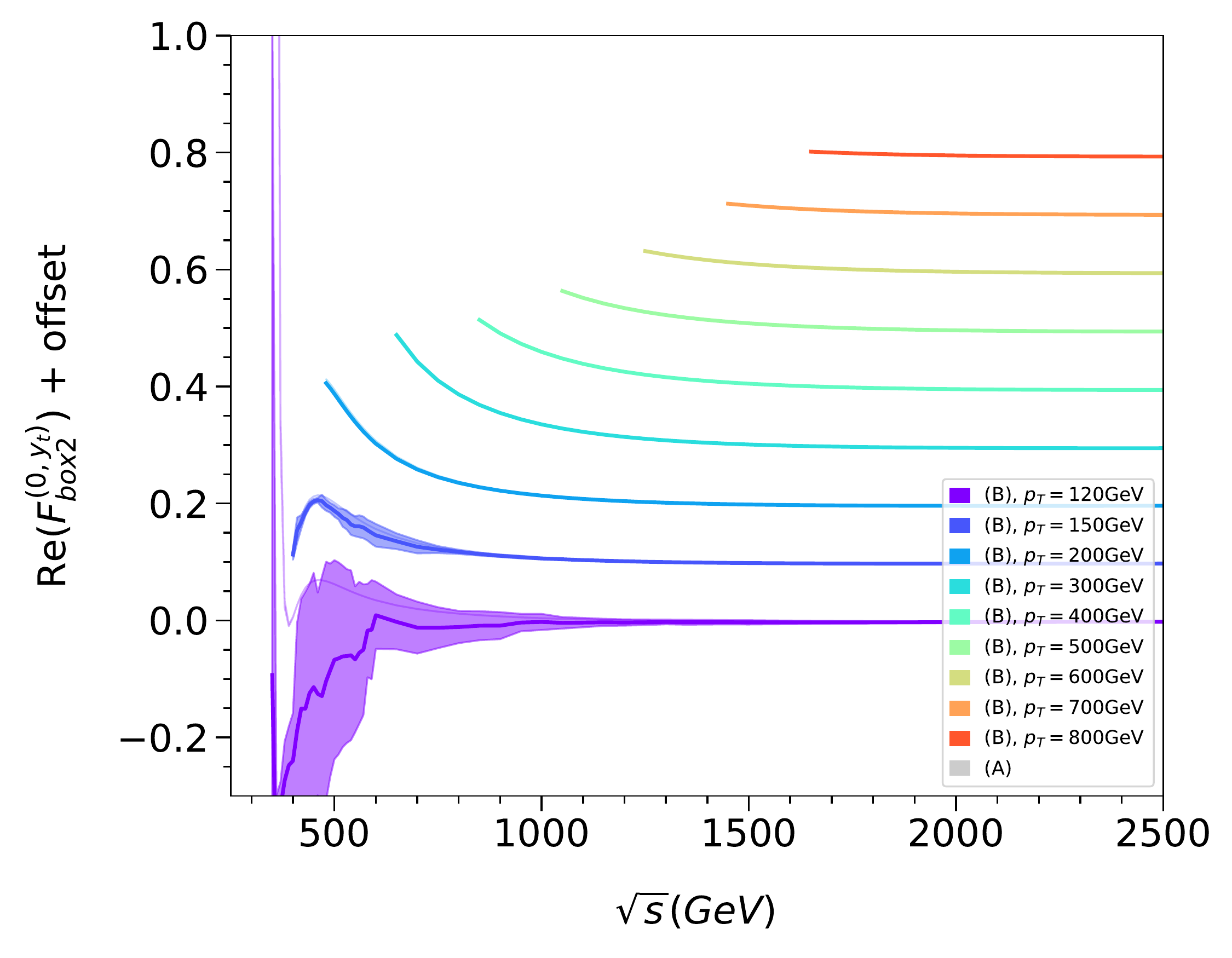}
    &
    \includegraphics[width=0.45\textwidth]{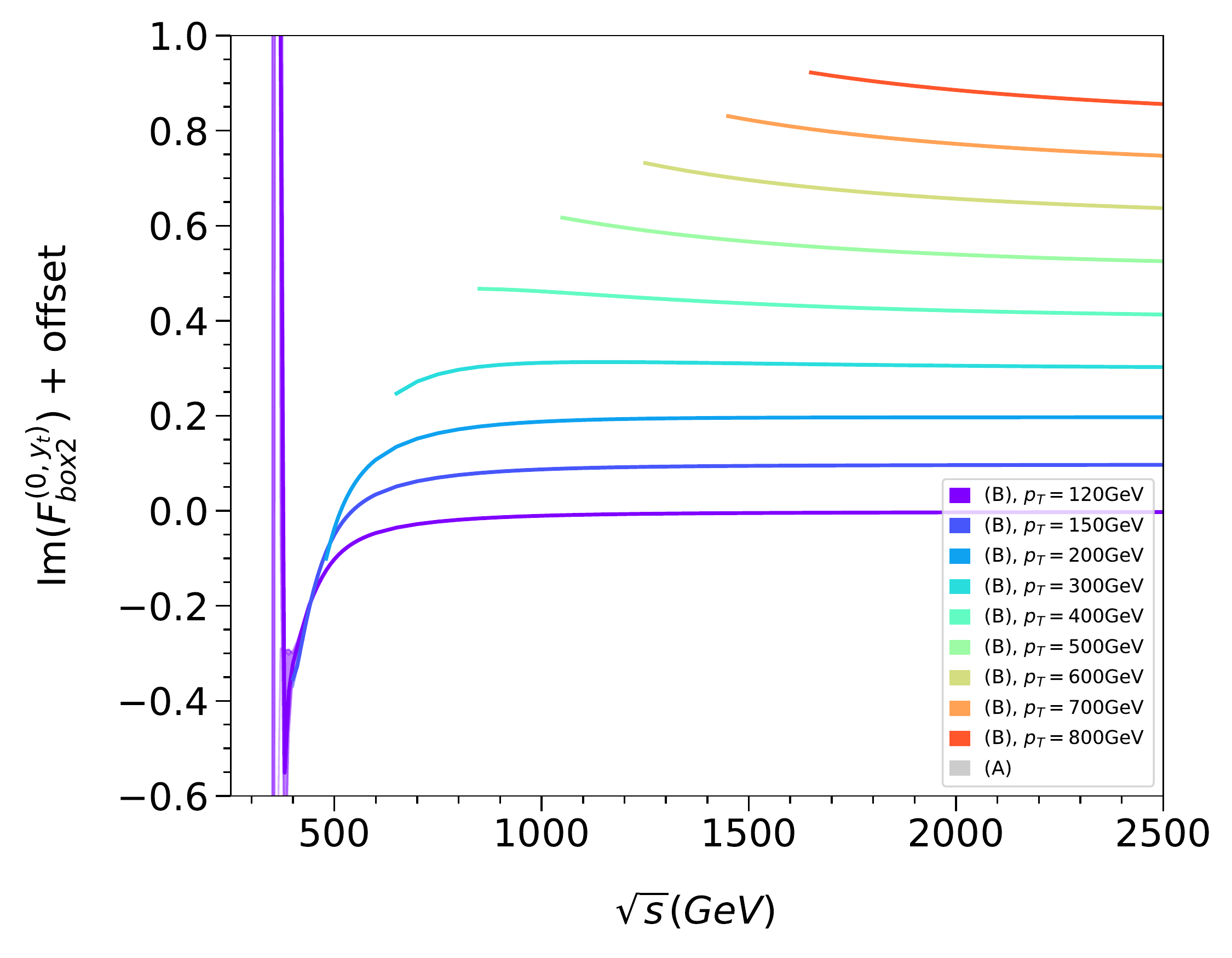}
  \end{tabular}
  \caption{\label{fig::F12}Real and imaginary parts of $F_{\rm box1}$ and
    $F_{\rm box2}$ for fixed $p_T$. Note that an offset is applied 
    such that the curves for the different $p_T$ values are clearly separated.
    No offset is used for the lowest $p_T$ value. For $F_1$ ($F_2$)
    we shift the subsequent $p_T$ curves by 1.5 (0.1) in the positive $y$ axis direction.
    The coloured curves and the corresponding bands correspond to 
    approach~(B). Results for approach~(A) are only shown as faint uncertainty
    bands.
    For $p_T\ge150$~GeV the central values of approach~(A) and~(B) agree.}
\end{figure}

Next we discuss the results for $F_{\rm box1}$ and $F_{\rm box2}$ for a range of
values for the transverse momentum $p_T$.  In Fig.~\ref{fig::F12} we show the real and
imaginary parts of $F_{\rm box1}$ and $F_{\rm box2}$ for $p_T$ between
$120$~GeV and $800$~GeV. 
The colours
correspond to the results from approach~(B); here we also show the uncertainty
band from the Pad\'e method.  The results from approach~(A) 
are shown as faint uncertainty bands. They are only visible for small
values of $p_T$, where one observes deviations between the two approaches.

Above $p_T=200$~GeV the uncertainty from the Pad\'e method is
negligible.  For $p_T=150$~GeV differences between the approaches are only
visible for the real part of $F_{\rm box2}$.  The situation is similar for
$p_T=120$~GeV for $\sqrt{s}\gsim 400$~GeV where the uncertainty bands are
still small.  Up to this value the results for $F_{\rm box1}$ and the
imaginary parts of $F_{\rm box2}$ agree quite well.  The real part of
$F_{\rm box2}$ shows larger uncertainties for large values of $\sqrt{s}$ in
approach~(B); for approach~(A), however, the uncertainties remain small.
Note, that $F_{\rm box2}$ is numerically less important than $F_{\rm box1}$.

Fig.~\ref{fig::F12} shows that both ways to treat the internal boson mass
leads (within uncertainties) to equivalent physical results. 
In view of the discussion above we expect that approach~(B) only approximates the
unknown exact result down to $\sqrt{s}\approx 520$~GeV. However,
approaches~(A) and~(B) agree for even smaller values of $\sqrt{s}$. It seems
that the master integrals of approach~(B) with non-analytic behaviour at the
three-particle threshold are numerically suppressed.

\begin{figure}[t]
  \centering
  \begin{tabular}{cc}
    \includegraphics[width=0.45\textwidth]{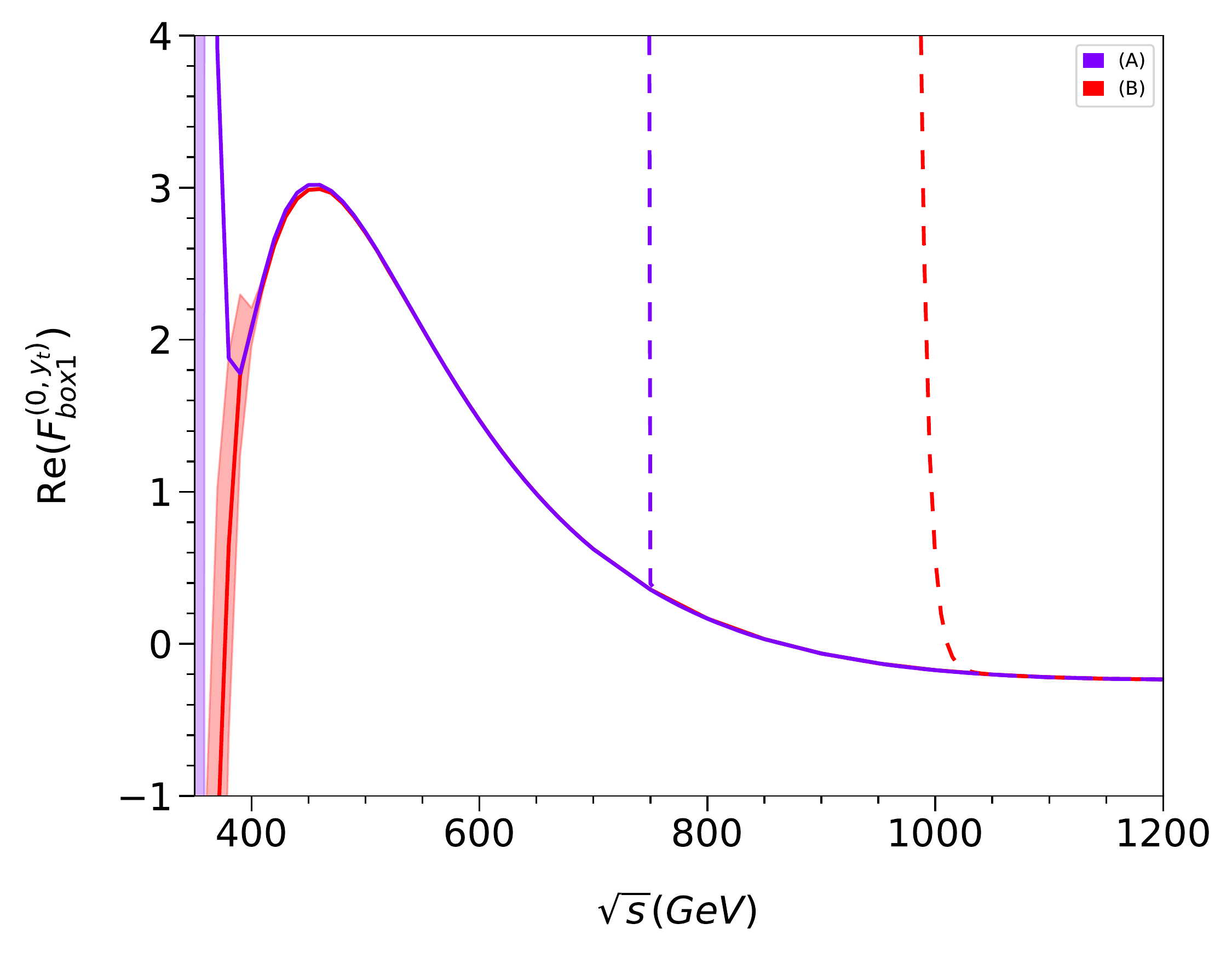}
    &
    \includegraphics[width=0.45\textwidth]{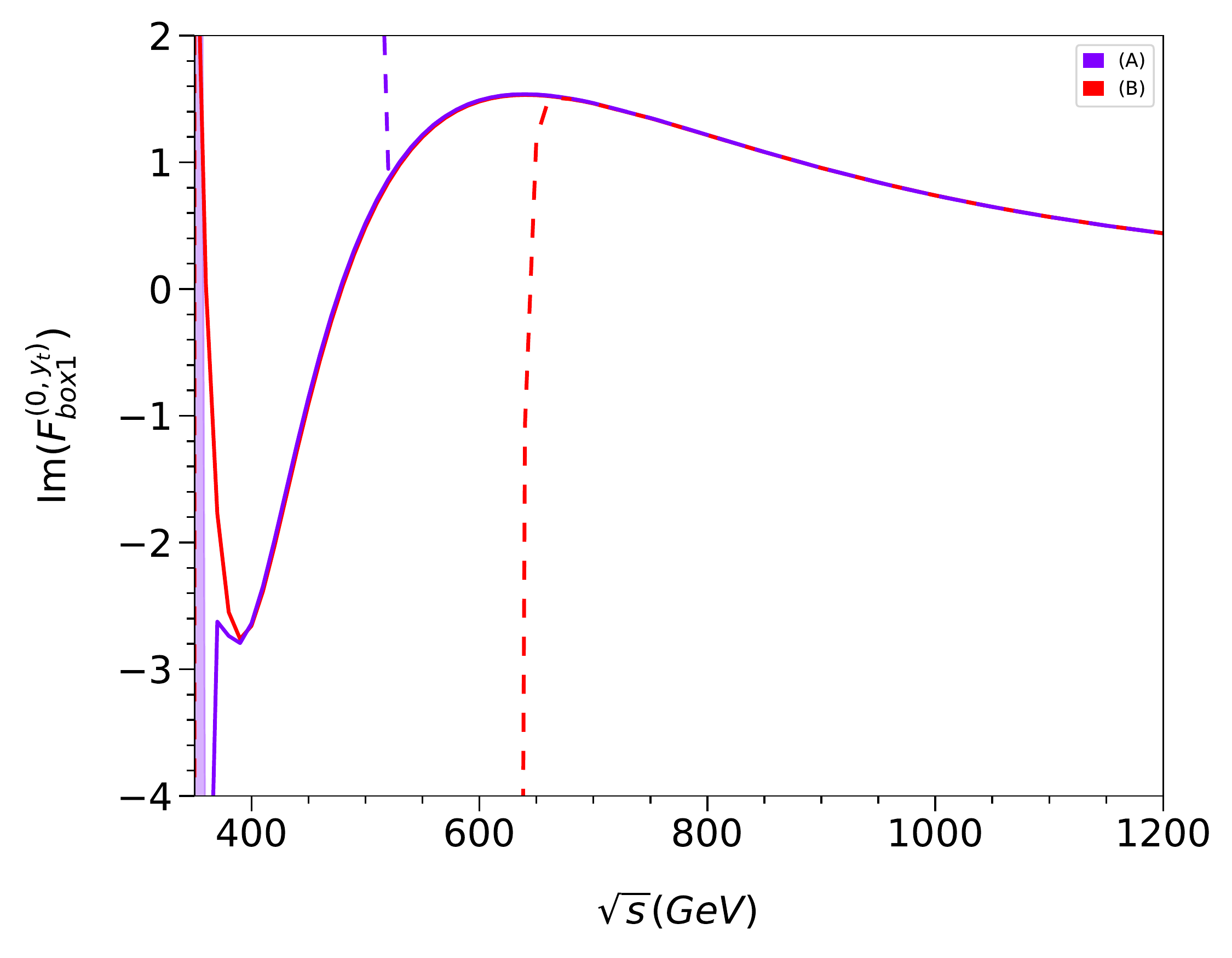}
    \\
    \includegraphics[width=0.45\textwidth]{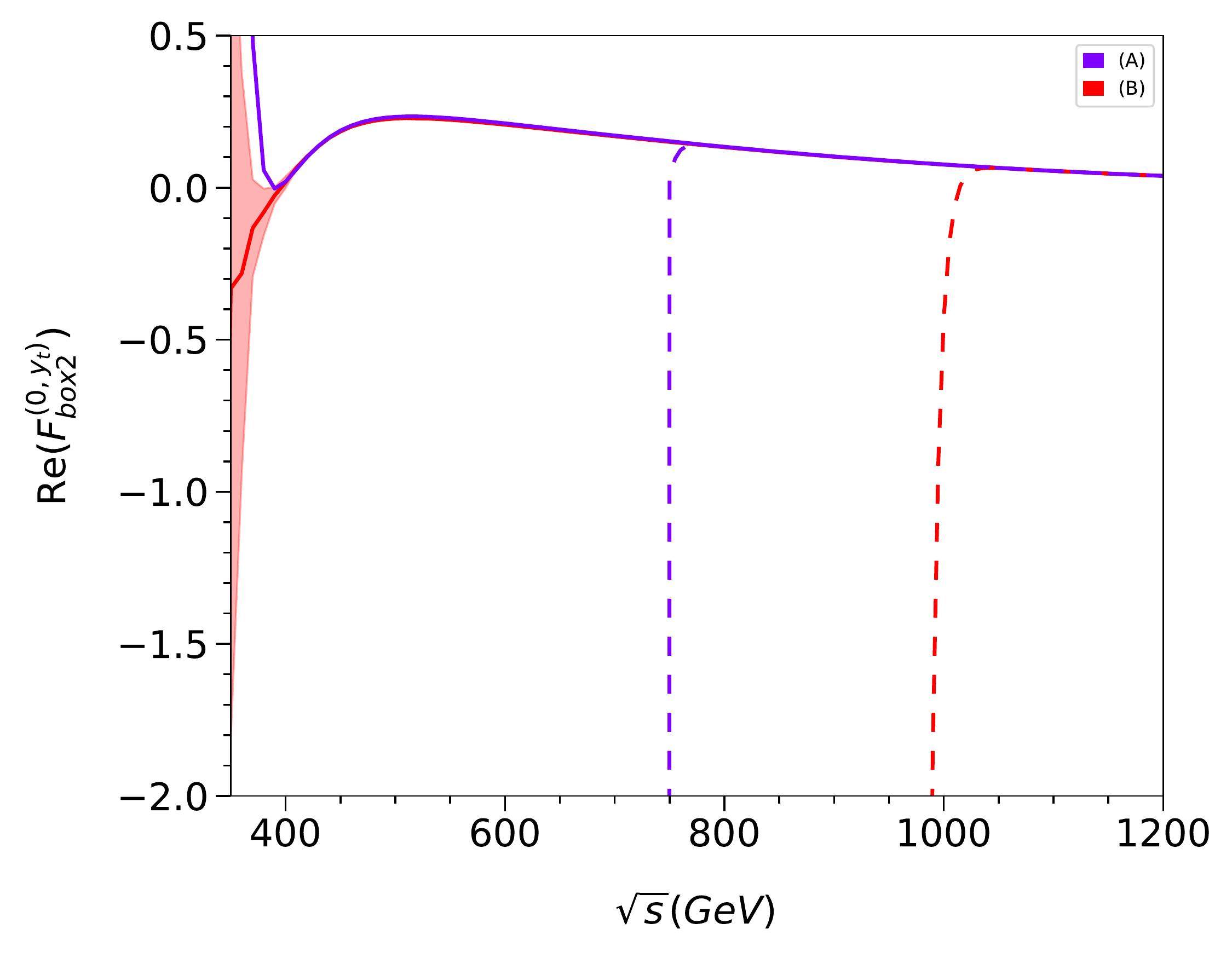}
    &
    \includegraphics[width=0.45\textwidth]{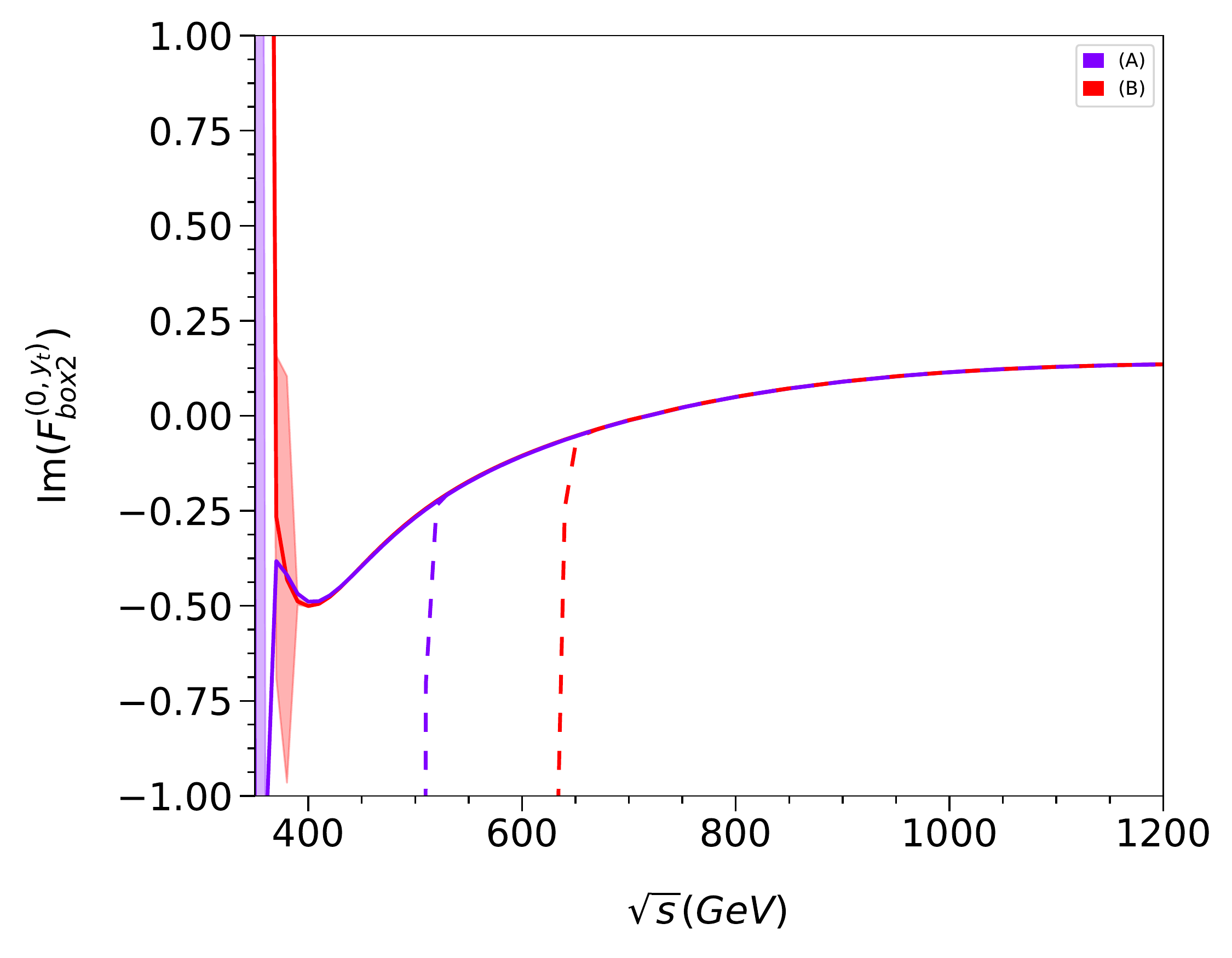}
  \end{tabular}
  \caption{\label{fig::F12_theta}Real and imaginary parts of $F_{\rm box1}$
    and $F_{\rm box2}$ for fixed scattering angle $\theta=\pi/2$.}
\end{figure}

In Fig.~\ref{fig::F12_theta} we show the real and imaginary parts of
$F_{\rm box1}$ and $F_{\rm box2}$ for fixed scattering angle $\theta=\pi/2$
for $\sqrt{s}$ between the top quark threshold and 1200~GeV. The solid curves
represent Pad\'e results and the dashed curves the expansions.  We
observe that the expansions start to diverge\footnote{In order to not
  overload the plots we only show the highest expansion terms in
  Fig.~\ref{fig::F12_theta}. The lower-order expansions show a similar
  behaviour.} for the real parts for $\sqrt{s}\approx 800$~GeV for
approach~(A) and for $\sqrt{s}\approx 1000$~GeV for approach~(B).  For the
imaginary parts the numbers are 700~GeV and 600~GeV, respectively.
Note, however, that the Pad\'e results of approaches~(A) and~(B)
are stable to fairly small values of $\sqrt{s}$. For $\sqrt{s}\gsim500$~GeV
the uncertainty bands are small and the two curves are on top of each
other. For smaller values of $\sqrt{s}$ the uncertainty band of approach~(B)
becomes bigger whereas the ones of approach~(A) remain small
in accordance with the discussion of the three-particle threshold at the
beginning of this section.

Let us finally perform a rough estimate of the numerical relevance of the
contributions computed in this paper. For this purpose we only compare the
real part of $F^{(0,y_t)}_{\rm box1}$ to the corresponding contribution from the
QCD corrections, $F^{(1,0)}_{\rm box1}$.  From Ref.~\cite{Davies:2018qvx} (see
also Section~3.3 of Ref.~\cite{Wellmann:2020rxl})
we find that $F^{(1,0)}_{\rm box1}$ is about ${\cal O}(1)$ if the scattering
angle is fixed to $\theta=\pi/2$ and for $\sqrt{s}$ a few hundred GeV.  This
is also the case for $F^{(0,y_t)}_{\rm box1}$ as can be seen from
Fig.~\ref{fig::F12_theta}. For the pre-factors in Eq.~(\ref{eq::F}) we have
$\alpha_t/\alpha_s \approx 0.6$ and thus it might very well be that the
electroweak corrections provide sizeable contributions to the Higgs pair cross
section. Of course, we should emphasize that in this paper only a certain
diagram class has been considered; in particular, no triangle diagrams are
included. Furthermore for this estimate no interference contributions are taken
into account.


\section{\label{sec::concl}Conclusions}

In this paper we take the first step towards the electroweak corrections to
Higgs boson pair production. We consider the subset of diagrams where a Higgs
boson is exchanged between the top quarks. Effects from Higgs self couplings
are neglected.

We are interested in analytic calculations of the form factors in the
high-energy limit; we perform expansions in $m_t^2/s$, $m_t^2/t$ and
$m_t^2/u$ taking into account up to about 60 expansion
terms. We study two methods for the treatment of the internal massive Higgs
boson, which is a new feature as compared to the QCD corrections. In our first
approach we assume that it is small as compared to the top quark mass, whereas
in the second approach it is assumed that the internal Higgs boson is of the
same order of magnitude as the top quark mass. In both cases we perform
expansions in the respective small parameters.  For physical values of the
mass parameters both expansion methods agree at the percent level
for smaller values of $p_T$ and at the permille level for larger values.

The approach with a small internal Higgs boson leads to master integrals
which have been computed in the context of QCD corrections. The other approach
leads to 140 new master integrals. We describe in detail our approach to
compute them analytically using differential equations and the Mellin Barnes
method. 

We supplement the expansion for small $m_t$ by combinations of Pad\'e approximations and the
associated uncertainty estimates, which significantly increases the region of phase space
where the analytic expansions can be used. We show that Pad\'e approximants
based on up to about 60 $m_t^2$ expansion terms provide excellent result
down to $p_T=150$~GeV and even for $p_T=120$~GeV results with moderate
uncertainties are obtained. On the basis of a scalar (master) integral we
validate that the uncertainty estimate covers the exact result.

The methods discussed in this paper are not restricted to internal Higgs
bosons. They can also be applied to internal gauge bosons and to other $2\to2$
processes mediated by a top quark loop and small external masses.  For the
subset of Feynman diagrams considered here only planar integrals contribute.  The
generalization to non-planar diagrams will be a challenge, however, we are
optimistic that they can be treated using the methods developed in this paper.



\section*{Acknowledgements}

We thank Gudrun Heinrich for comments on the draft.
This research was supported by the Deutsche Forschungsgemeinschaft (DFG,
German Research Foundation) under grant 396021762 --- TRR 257
``Particle Physics Phenomenology after the Higgs Discovery''.
The work of G.M. was supported by JSPS KAKENHI (No. JP20J00328).
The work of J.D.~was in part supported by the Science and Technology
Facilities Council (STFC) under the Consolidated Grant ST/T00102X/1.


\appendix


\section{Constants from three-dimensional MB integrals}

\label{app:MBconst}
The three-dimensional MB representations and the analytic expressions for the constants $C_T$ and $C_S$ present in the
``7+2''-line integral are given by
{\scalefont{0.8}
\begin{eqnarray}
C_{T} \!\!\!&=\!\!&
\int_{z_1 z_2 z_3} \!\!\!\!\! -\frac{2 \Gamma\left[z_{12}+2,z_{12}+2,-z_3,z_3+1,z_3-z_1,z_{23}+2,-z_1,z_1+1,-z_2,-z_{23}-2, z_{123}+3\right]}{\Gamma\left[z_1+2,z_{12}+3,z_2+2 z_3+3\right]} \nonumber \\[1mm]
&& -\, \frac{4 \Gamma\left[z_{12}+2,-z_{23}-1,-z_3,z_3+1,z_3-z_1,z_{23}+2,-z_1,z_1+1,-z_2,z_{12}+1, z_{123}+3 \right]}{\Gamma\left[z_1+2,z_{12}+3,z_2+2 z_3+3\right]} \nonumber \\[1mm]
&&-\, \frac{4 \Gamma\left[z_{12}+2,-z_{23}-1,-z_3,z_3+2,z_3-z_1,z_{23}+1,-z_1,z_1+1,-z_2,z_{12}+1, z_{123}+3 \right]}{\Gamma\left[z_1+2,z_{12}+3,z_2+2 z_3+3\right]} \nonumber \\[1mm]
&&-\, \frac{2 \Gamma\left[z_{12}+2,z_{12}+2,-z_3,z_3+1,-z_1+z_3+1,z_{23}+3,-z_1,z_1+1,-z_2,-z_{23}-3, z_{123}+3\right]}{\Gamma\left[z_1+2,z_{12}+3,z_2+2 z_3+4\right]}\nonumber\\[1mm]
&&-\, \frac{2 \Gamma\left[z_{12}+2,-z_{23}-2,-z_3,z_3+1,-z_1+z_3+1,z_{23}+3,-z_1,z_1+1,-z_2,z_{12}+1, z_{123}+3\right] }{\Gamma\left[z_1+2,z_{12}+3,z_2+2 z_3+4\right]} \nonumber\\[1mm]
&&-\,\frac{6 \Gamma\left[z_{12}+2,-z_{23}-2,-z_3,z_3+2,-z_1+z_3+1,z_{23}+2,-z_1,z_1+1,-z_2,z_{12}+1,z_{123}+3\right]}{\Gamma\left[z_1+2,z_{12}+3,z_2+2 z_3+4\right]} \nonumber \\[1mm]
&&-\, \frac{2 \Gamma\left[z_{12}+2,-z_{23}-3,-z_3,z_3+2,-z_1+z_3+2,z_{23}+3,-z_1,z_1+1,-z_2,z_{12}+1,z_{123}+3\right]}{\Gamma\left[z_1+2,z_{12}+3,z_2+2 z_3+5\right]} \,, \nonumber \\
&=& \frac{5}{3} + \frac{\pi^2}{18} + \frac{88\pi^4}{405}
    - 8 \zeta_3
    - \frac{8\pi^2}{27} \psi ^{(1)}\left(\tfrac{1}{3}\right)
    + \frac{2}{9}  \left[ \psi ^{(1)}\left(\tfrac{1}{3}\right)\right]^2
    \nonumber\\
&\approx& 6.890254528\ldots \,,
\end{eqnarray}
}
and
{\scalefont{0.8}
\begin{eqnarray}
C_{S}  \!\!\!&=\!\!&
\int_{z_1 z_2 z_3} \!\!\!\!\!
\frac{2 \Gamma\left[ z_{12}+2,z_{12}+2,-z_3,z_3+1,-z_1+z_3-1,z_{23}+1,-z_1,z_1+1,-z_2,-z_{23}-1,z_{123}+3\right]}{\Gamma\left[z_1+2,z_{12}+3,z_2+2 z_3+2\right]} \nonumber \\[1mm]
&&-\, \frac{2 \Gamma \left[ z_{12}+2,-z_{23}-1,-z_3,z_3+1,z_3-z_1,z_{23}+2,-z_1,z_1+1,-z_2,z_{12}+1, z_{123}+3\right]}{\Gamma\left[ z_1+2,z_{12}+3,z_2+2 z_3+3\right]} \nonumber \\[1mm]
&&-\,\frac{2 \Gamma\left[z_{12}+2,-z_{23}-1,-z_3,z_3+2,z_3-z_1,z_{23}+1,-z_1,z_1+1,-z_2,z_{12}+1, z_{123}+3\right]}{\Gamma\left[z_1+2,z_{12}+3,z_2+2 z_3+3\right]} \nonumber \\[1mm]
&&- \, \frac{2 \Gamma\left[z_{12}+2,z_{12}+2,-z_3,z_3+1,-z_1+z_3+1,z_{23}+3,-z_1,z_1+1,-z_2,-z_{23}-3, z_{123}+3 \right]}{\Gamma\left[z_1+2,z_{12}+3,z_2+2 z_3+4\right]}\nonumber \\[1mm]
&&-\, \frac{2 \Gamma\left[z_{12}+2,-z_{23}-2,-z_3,z_3+1,-z_1+z_3+1,z_{23}+3,-z_1,z_1+1,-z_2,z_{12}+1,z_{123}+3 \right]}{\Gamma\left[z_1+2,z_{12}+3,z_2+2 z_3+4\right]} \nonumber \\[1mm]
&&-\,\frac{4 \Gamma\left[z_{12}+2,-z_{23}-2,-z_3,z_3+2,-z_1+z_3+1,z_{23}+2,-z_1,z_1+1,-z_2,z_{12}+1,z_{123}+3 \right]}{\Gamma\left[z_1+2,z_{12}+3,z_2+2 z_3+4\right]}\nonumber \\[1mm]
&&-\,\frac{2
   \Gamma\left[z_{12}+2,-z_{23}-3,-z_3,z_3+2,-z_1+z_3+2,z_{23}+3,-z_1,z_1+1,-z_2,z_{12}+1, z_{123}+3\right]}{\Gamma\left[z_1+2,z_{12}+3,z_2+2 z_3+5\right]} \,, \nonumber \\
&=& \frac{2}{3} - \frac{5\pi^2}{18} + \frac{649\pi^4}{1620}
    - 12 \zeta_3
    - \frac{20\pi^2}{27} \psi ^{(1)}\left(\tfrac{1}{3}\right)
    + \frac{5}{9}  \left[ \psi ^{(1)}\left(\tfrac{1}{3}\right)\right]^2
    \nonumber\\
&\approx& 5.339941546 \ldots  \,,
\end{eqnarray}            
}where the integration contours are fixed at $\{ \mathrm{Re}(z_1) = -1/7, \mathrm{Re}(z_2) = -1/11, \mathrm{Re}(z_3) = -1/17 \}$.

The analytic result for $C_T$ is obtained from a consistency condition
obtained from the system of $m_t$-expanded $t$-differential equations for the
140 master integrals.  On the other hand, for $C_S$ we first perform various
shifts of integration contours and analytic continuations to bring the
three-dimensional MB integrals into a better form, which can be reduced to,
at most, two-dimensional integrals in terms of only Gamma functions by the Barnes
lemmas. The resulting MB integrals are the solved by the analytical
summations and \texttt{PSLQ} algorithm.
Note that it is straightforward to directly compute $C_S$ and $C_T$ numerically and obtain
a precision of about ten digits, which is sufficient for practical
applications.





\end{document}